\documentclass[12pt]{article}
 \pdfoutput=1
\usepackage[table]{xcolor}
\usepackage{amsmath,amssymb,exscale}
\usepackage{graphicx}
\usepackage{epsfig}
\usepackage{multicol}
\usepackage{color}
\usepackage{mathrsfs}
\usepackage{blindtext}
 \usepackage{fancyhdr}
\usepackage{hyperref}
\usepackage{cite}
\usepackage{mathtools}
\usepackage{amsmath}
\usepackage{rotating,slashed,amsmath,charter,xcolor,catchfilebetweentags,ifluatex}

\usepackage{graphicx}
\usepackage{sidecap}

\usepackage[latin1]{inputenc} 
\usepackage{multicol}  
 
 \usepackage{titlesec}
 
 \usepackage{rotating,slashed,xcolor,amsfonts,expdlist,charter}

\numberwithin{equation}{section}

\usepackage{xcolor}
\usepackage{sectsty}

\usepackage{mdframed}
\usepackage{titletoc}

\definecolor{secnum}{RGB}{13,151,225}
\definecolor{ptcbackground}{RGB}{212,237,252}
\definecolor{ptctitle}{RGB}{0,177,235}

\titlecontents{lsection}
  [5.8em]{\sffamily}
  {\color{secnum}\contentslabel{2.3em}\normalcolor}{}
  {\titlerule*[1000pc]{.}\contentspage\\\hspace*{-5.8em}\vspace*{5pt}%
    \color{white}\rule{\dimexpr\textwidth-15.5pt\relax}{1pt}}

\usepackage{hyperref}
\hypersetup{colorlinks,bookmarksopen,bookmarksnumbered,citecolor=blus,
linkcolor=redy,pdfstartview=FitH,urlcolor=blus}
\usepackage{slashed}

\definecolor{blus}{cmyk}{1,0.9,0,0.1}
\definecolor{verdes}{cmyk}{0.99,0,0.59,0.65}
\definecolor{rossos}{cmyk}{0,1,1,0.55}
\definecolor{redy}{cmyk}{0,1,1,0.7}
\definecolor{greeny}{cmyk}{0.99,0,0.59,0.98}
\definecolor{green-go}{cmyk}{0.79,0,0.59,0.5}

\usepackage{titlesec}

\newcommand{\beq}{\begin{equation}}
\newcommand{\eeq}{\end{equation}}

\def\hhref#1{\href{http://arxiv.org/abs/#1}{arXiv:#1}} 

\newcommand{\tmtextbf}[1]{{\bfseries{#1}}}
\newcommand{\tmtextrm}[1]{{\rmfamily{#1}}}
\def\bp{M_{P}}

\def\be{\begin{equation}}
\def\ee{\end{equation}}
\def\ba{\begin{array} }

\def\bac{\begin{array} {c}}
\def\bacc{\begin{array} {cc}}
\def\baccc{\begin{array} {ccc}}
\def\bacccc{\begin{array} {cccc}}
\def\ea{\end{array}}
\def\bea{\begin{eqnarray}}
\def\eea{\end{eqnarray}}

\definecolor{red}{rgb}{1,0,0}

\def\psl{\hbox{\hbox{${p}$}}\kern-1.9mm{\hbox{${/}$}}}
\def\dsl{\hbox{\hbox{${\partial}$}}\kern-2.2mm{\hbox{${/}$}}}
\def\Dsl{\hbox{\hbox{${D}$}}\kern-2.6mm{\hbox{${/}$}}}

\newcommand{\gappeq}{{\rlap{{\raise}.5ex\text{\ensuremath{>}}}{{\lower}.5ex\text{\ensuremath{\sim}}}}}
\newcommand{\lappeq}{{\rlap{{\raise}.5ex\text{\ensuremath{<}}}{{\lower}.5ex\text{\ensuremath{\sim}}}}} 
\newcommand{\I}{\tmtextrm{1{\kern}-.24em l}}

\begin{document}
\topmargin -1.0cm
\oddsidemargin 0.9cm
\evensidemargin -0.5cm

{\vspace{-1cm}}
\begin{center}

\vspace{-1cm}

 {\huge \tmtextbf{ 
\color{rossos} (In)equivalence of Metric-Affine and Metric Effective Field Theories}} {\vspace{.5cm}}\\

\vspace{1.9cm}

{\large  {\bf Gianfranco Pradisi} and {\bf Alberto Salvio }
{\em  

\vspace{.4cm}

 Physics Department, University of Rome Tor Vergata, \\ 
via della Ricerca Scientifica, I-00133 Rome, Italy\\

\vspace{0.6cm}

I. N. F. N. -  Rome Tor Vergata,\\
via della Ricerca Scientifica, I-00133 Rome, Italy\\ 

\vspace{0.4cm}

\vspace{0.2cm}

 \vspace{0.5cm}
}

\vspace{0.2cm}

}
\vspace{0.cm}

%
%
 %
%
%

\end{center}

%
 
\noindent --------------------------------------------------------------------------------------------------------------------------------

\begin{center}
{\bf \large Abstract}
\end{center}

\noindent 
 In a geometrical approach to gravity the metric and the (gravitational) connection can be independent and one deals with metric-affine  theories.   
We construct the most general action of metric-affine effective field theories, including a generic matter sector, 
where the connection does not carry additional dynamical fields. 
Among other things, this helps in identifying the complement set of effective field theories where there are other dynamical fields, which can have an interesting phenomenology. 
 Within the latter set, we study in detail a vast class where the Holst invariant (the contraction of the curvature with the Levi-Civita antisymmetric tensor) is a dynamical pseudoscalar. 
 In the Einstein-Cartan case (where the connection is metric compatible and fermions can be introduced) we also comment on the possible phenomenological role of dynamical dark photons from torsion and compute interactions of the above-mentioned pseudoscalar with a generic matter sector and the metric.
Finally, we show that in an arbitrary realistic metric-affine theory featuring a generic matter sector the equivalence principle always emerges at low energies without the need to postulate it.

  \vspace{0.4cm}

\noindent --------------------------------------------------------------------------------------------------------------------------------

\vspace{1.1cm}

\vspace{2cm}

\tableofcontents

\vspace{0.5cm}

\section{Introduction}\label{Introduction}

General relativity (GR) is an extremely successful theory of gravity, which agrees with all observations performed so far. Recent tests of GR include the  discovery of gravitational waves, whose production is consistent with coalescing black holes~\cite{LIGOScientific:2016aoc}, and the images of the black holes in the center of the M87 and our galaxy produced by the Event Horizon Telescope~\cite{EventHorizonTelescope:2019dse,EventHorizonTelescope:2019ggy,EventHorizonTelescope:2022xnr,EventHorizonTelescope:2022wok,EventHorizonTelescope:2022xqj}. 

Of course, GR has to be complemented by some matter fields. At least a set of spin-1, spin-1/2 and spin-0 fields are needed to describe all we know about non-gravitational physics, the Standard Model of particles (SM) and its extensions  that can account for the evidence of beyond-the-SM physics (neutrino masses and mixings, dark matter, baryon asymmetry, etc.).

Moreover, a UV completion is also necessary because GR is known to be nonrenormalizable by perturbative methods~\cite{Goroff:1985sz,Goroff:1985th} and to be within the regime of validity of perturbation theory at energies much below the Planck scale. 
However, at those low energies we can construct a consistent theory  by adding all possible  operators along the lines of effective field theories~\cite{Donoghue:1994dn} (see also Refs.~\cite{Burgess:2003jk,Burgess:2007pt} for reviews). The lower the dimensionality of a given operator is the more relevant such operator is expected to be at low energies.

The main principle behind these constructions, including GR itself, is general covariance (or the general relativity principle), which essentially states that all laws of physics should be invariant under a general coordinate transformation. This also implies that the field equations can be written in a covariant form and renders the presence of tensors, such as the metric, and a connection necessary. This geometrization of physics is commonly regarded as one of the greatest achievements of Einstein's theory.   In GR and its effective field theory (EFT) extensions, including ordinary matter fields (spin-1, spin-1/2 and spin-0 fields), the connection is typically assumed to be the Levi-Civita one, a functional of the metric. Theories of this sort are thus called metric theories. However, from the geometrical point of view the metric and the connection can be completely independent objects.

Therefore, a natural modification of gravity can be obtained by promoting the connection to an independent degree of freedom, but preserving general covariance. The resulting theories are called metric-affine (see Ref.~\cite{Baldazzi:2021kaf} for a recent discussion on this subject and references to other original articles and Ref.~\cite{Hehl:1994ue} for a classic review). In general the difference between an arbitrary connection and the Levi-Civita connection is a tensor, known as the distorsion. The distorsion coincides with the contorsion when the theory is metric compatible, i.e.~when the covariant derivate of the metric vanishes, which is required by the presence of fermions. The contorsion in turn is a tensor that can be expressed in terms of the torsion and that vanishes if and only if the torsion does. The metric-compatible theories are also known as Poincar\'e gauge theories because they can always be formulated as theories with a local Poincar\'e symmetry 
(see~\cite{Obukhov:2022khx} for a recent review with many references to original works).

One of the purposes of the present paper is to identify the general\footnote{For previous less general studies see Refs.~\cite{Vitagliano:2010sr,Vitagliano:2013rna,Karananas:2021zkl}.} form of the action of a {\it metric-affine} EFT that is equivalent to a {\it metric} EFT in the sense that does not feature an independent dynamical distorsion: i.e. the distorsion can be exactly integrated out and expressed in terms of the metric and the matter   fields that are {\it not} of gravitational origin (that do not come from the metric and/or the distorsion). Indeed, even in a metric EFT additional gravitational degrees of freedom besides the massless spin-2 graviton can emerge from the metric because higher powers of the curvature tensors (that can involve higher derivatives) are generically present.

 The motivation for finding the general action described in the previous paragraph is the fact that it helps us to tell whether a given metric-affine theory does not feature an independent dynamical distorsion  without performing a direct calculation of the dynamical degrees of freedom. Also, with this result in hand, one could automatically identify the complement set of metric-affine EFTs that can potentially feature an independent dynamical distorsion. 
This set of theories is particularly interesting as the new distorsion fields can have interesting phenomenological consequences.

Another purpose of this paper is to discuss the validity of the equivalence principle in these EFTs. The equivalence principle is often presented as the starting point in formulating GR. However, in a metric EFT this principle  is generically broken by the higher-dimensional operators. Given that GR plus minimally-coupled matter fields  anyhow describe the low-energy limit of metric EFTs  the equivalence principle is always recovered at low energies in metric theories. It is then natural to ask whether the same is true in general {\it metric-affine} EFTs: is the equivalence principle always an emergent low energy property in an arbitrary theory?

 Let us now give an outline of the paper (a detailed summary of the results will be given in the concluding section). 
 In Sec.~\ref{Ingredients} we will present the key ingredients that are needed to construct metric-affine EFTs. We will not limit ourselves to the gravitational sector, but we will also include a general matter content, namely an arbitrary number of scalars (or pseudoscalars), gauge fields and fermions. 
The general action of theories with non-dynamical distorsion will then be the topic of Sec.~\ref{eq-theories}.  
After that, in Sec.~\ref{Dyn-Dist}, we will  discuss theories with dynamical distorsion, studying in detail some explicit examples. 
The possible breaking of the equivalence principle and its possible emergence at low energies in metric-affine theories will then be investigated in Sec.~\ref{A note on the equivalence principle}. Finally, in the concluding Sec.~\ref{Conclusions} we offer a detailed summary of the new results of the paper with some further discussions.

\section{Ingredients} \label{Ingredients}

 In this section we provide the main ingredients that are needed to construct  gravitational theories coupled to a generic matter sector. Most of the material in this section is a review of well-known results, but it is all needed to  understand the subsequent sections. Here we will also take advantage to fix our notation. 
 
 To describe gravity we start from the {\it general relativity principle}, which states that all laws of physics should be invariant under general coordinate transformations.
 To implement such principle  we introduce a metric $g_{\mu\nu}$ and a connection ${\cal A}_{\mu~\sigma}^{~\,\rho}$ as independent fields. So we are in the framework of metric-affine theories. 
 
 The metric would be needed even if gravity were absent, indeed writing the flat metric\footnote{$\eta_{ab}$ represents the Minkowski metric and is needed  to recover all we know about non-gravitational physics. We use the mostly plus signature convention $\{\eta_{ab}\}=$~diag$(-1,1,1,1)=\{\eta^{ab}\}.$} $ds^2=\eta_{ab}d\xi^ad\xi^b$ in general coordinates $x^\mu$ the metric $g_{\mu\nu}$ appears: $ds^2=g_{\mu\nu}dx^\mu dx^\nu$. The general transformation rule of the metric (obtained by requiring $ds^2$ invariant) is
 \be g'_{\alpha\beta}(x') = \frac{\partial x^\mu}{\partial x^{'\alpha}}\frac{\partial x^\nu}{\partial x^{'\beta}} g_{\mu\nu}(x) \label{trang}
 \ee 
 and, generically, in the presence of gravity it is not possible to recover the flat metric with a general coordinate transformation.
 
  The connection, on the other hand, is needed in curved space to introduce covariant derivatives of tensors, which are essential to write the field equations (which contain derivatives) in a covariant form: the covariant derivatives of a generic tensor $T_{\mu_1...\mu_n}^{\nu_1...\nu_m}$ with $n$ covariant indices and $m$ contravariant\footnote{As usual a covariant vector is an object that transforms as $\frac{\partial}{\partial x^\mu}$  and a contravariant vector is an object that transforms as $dx^\mu$ under general coordinate transformations. A tensor with $n$ covariant indices and $m$ contravariant indices transforms as the direct product of $n$ covariant vectors and $m$ contravariant vectors. Note that $g_{\mu\nu}$ is a tensor with two covariant indices (see Eq.~(\ref{trang})). The inverse metric $g^{\mu\nu}$, i.e. $g^{\mu\rho}g_{\rho\nu} = \delta^\mu_\nu$, (which exists, as shown below) is a tensor with two contravariant indices because $\delta^\mu_\nu$ is invariant. As usual here we raise and lower the spacetime indices through the inverse metric $g^{\mu\nu}$  and $g_{\mu\nu}$, respectively. The flat indices $a,b,...$ are raised and lowered with $\eta^{ab}$ and $\eta_{ab}$, respectively.} indices are
\be {\cal D}_\mu T_{\mu_1...\mu_n}^{\nu_1...\nu_m} =\partial_\mu T_{\mu_1...\mu_n}^{\nu_1...\nu_m} +{\cal A}_{\mu~\beta_1}^{~\,\nu_1}T_{\mu_1...\mu_n}^{\beta_1...\nu_m} +...+{\cal A}_{\mu~\beta_m}^{~\,\nu_m}T_{\mu_1...\mu_n}^{\nu_1...\beta_m}-{\cal A}_{\mu~\mu_1}^{~\,\alpha_1}T_{\alpha_1...\mu_n}^{\nu_1...\nu_m} -...-{\cal A}_{\mu~\mu_n}^{~\,\alpha_n}T_{\mu_1...\alpha_n}^{\nu_1...\nu_m}.\label{CovDer}\ee
This calligraphic covariant derivative ${\cal D}$  is generically different from the covariant derivative, which we denote $D$, computed with the Levi-Civita (LC) connection 
\be \Gamma_{\mu~\sigma}^{~\,\rho} = \frac12 g^{\rho\tau}\left(\partial_\mu g_{\tau\sigma}+\partial_\sigma g_{\tau\mu}-\partial_\tau g_{\mu\sigma}\right).\label{LCC}\ee
In order for the quantity in~(\ref{CovDer}) to  be a tensor with $m$ contravariant indices and $n+1$ covariant indices ${\cal A}_{\mu~\sigma}^{~\,\rho}$ should transform under general coordinate transformations precisely as $\Gamma_{\mu~\sigma}^{~\,\rho}$. So 
\be C_{\mu~\sigma}^{~\,\rho} \equiv {\cal A}_{\mu~\sigma}^{~\,\rho}-\Gamma_{\mu~\sigma}^{~\,\rho}, \ee
which we call the distorsion,
transforms as a tensor.  Theories where $C_{\mu~\sigma}^{~\,\rho}=0$ are called metric theories as the connection can be computed once the metric is known in that case. The  torsion $T_{\mu\nu\rho}$ is defined in terms of the distorsion by
 \be T_{\mu\nu\rho} \equiv C_{\mu\nu\rho}-C_{\rho\nu\mu}, \label{torsion-distorsion}\ee 
which is antisymmetric with respect to the exchange $\mu\leftrightarrow\rho$.
The curvature associated with ${\cal A}_{\mu~\sigma}^{~\,\rho}$ is defined by 
\be {\cal F}_{\mu\nu~~\sigma}^{~~~\rho} \equiv \partial_\mu{\cal A}_{\nu~\sigma}^{~\,\rho}-\partial_\nu{\cal A}_{\mu~\sigma}^{~\,\rho}+{\cal A}_{\mu~\lambda}^{~\,\rho}{\cal A}_{\nu~\sigma}^{~\,\lambda}-{\cal A}_{\nu~\lambda}^{~\,\rho}{\cal A}_{\mu~\sigma}^{~\,\lambda},\ee
which can be expressed in terms of $C_{\mu~\sigma}^{~\,\rho}$ as 
\be {\cal F}_{\mu\nu~~\sigma}^{~~~\rho}=R_{\mu\nu~~\sigma}^{~~~\rho} +D_\mu C_{\nu~\sigma}^{~\,\rho}-D_\nu C_{\mu~\sigma}^{~\,\rho}+ C_{\mu~\lambda}^{~\,\rho} C_{\nu~\sigma}^{~\,\lambda}- C_{\nu~\lambda}^{~\,\rho} C_{\mu~\sigma}^{~\,\lambda},  \label{FRC}  \ee
where $R_{\mu\nu~~\sigma}^{~~~\rho}$ is the standard Riemann tensor\footnote{We use the conventions
$$R_{\mu\nu\,\,\, \sigma}^{\quad \rho} \equiv \partial_{\mu} \Gamma_{\nu \, \sigma}^{\,\rho}- \partial_{\nu} \Gamma_{\mu \, \sigma}^{\,\rho} +  \Gamma_{\mu \, \tau}^{\,\rho}\Gamma_{\nu \,\sigma }^{\,\tau}- \Gamma_{\nu \, \tau}^{\,\rho}\Gamma_{\mu \,\sigma }^{\,\tau}, \quad R_{\mu\nu} \equiv R_{\rho\mu\,\,\, \nu}^{\quad \rho},\quad R\equiv g^{\mu\nu}R_{\mu\nu}.$$}, i.e. ${\cal F}_{\mu\nu~~\sigma}^{~~~\rho}$ evaluated at ${\cal A}_{\mu~\sigma}^{~\,\rho}=\Gamma_{\mu~\sigma}^{~\,\rho}$. Starting from  ${\cal F}_{\mu\nu~~\sigma}^{~~~\rho}$ we can define a scalar 
\be {\cal R} \equiv {\cal F}_{\mu\nu}^{~~~\mu\nu}\ee
and a pseudoscalar (see \cite{Hojman:1980kv,Nelson:1980ph,Holst:1995pc})
\be {\cal R'} \equiv \frac1{\sqrt{-g}}\epsilon^{\mu\nu\rho\sigma}{\cal F}_{\mu\nu\rho\sigma},\ee
where $\epsilon^{\mu\nu\rho\sigma}$ is the totally antisymmetric Levi-Civita symbol with $\epsilon^{0123}=1$. We will refer to ${\cal R'}$ as the Holst invariant. The pseudoscalar ${\cal R'}$ vanishes for $C_{\mu~\sigma}^{~\,\rho}=0$ (that is when the connection is the LC one) because of the cyclicity property $R_{\mu\nu\rho\sigma}+R_{\nu\sigma\rho\mu}+R_{\sigma\mu\rho\nu}=0$, which is the reason why in standard Riemannian geometry ${\cal R'}$ is absent. Therefore, ${\cal R'}$ can be considered as a direct manifestation of a connection that is independent of the metric. We will study its possible dynamics in Sec.~\ref{(pseudo)scalaron}. By using~(\ref{FRC}) one obtains
\bea {\cal R} &=& R +D_\mu C_{\nu}^{~\,\mu\nu}-D_\nu C_{\mu}^{~\,\mu\nu}+ C_{\mu~\lambda}^{~\,\mu} C_{\nu}^{~\,\lambda\nu}- C_{\nu~\lambda}^{~\,\mu} C_{\mu}^{~\,\lambda\nu}, \label{RRC} \\
{\cal R'} &=& \frac{2}{\sqrt{-g}}\epsilon^{\mu\nu\rho\sigma}\left(D_\mu C_{\nu\rho\sigma}+C_{\mu\rho\lambda} C_{\nu~\sigma}^{~\,\lambda}\right). \label{RpRC}\eea
 Note that we can decompose 
\be {\cal F}_{\mu\nu\rho\sigma} = \frac1{16} g_{\mu\rho}g_{\nu\sigma} {\cal R} -\frac1{4!\sqrt{-g}}\epsilon_{\mu\nu\rho\sigma}{\cal R}'+\tilde {\cal F}_{\mu\nu\rho\sigma}, \qquad(\tilde{\cal F}_{\mu\nu}^{~~~\mu\nu}=0,~~\epsilon^{\mu\nu\rho\sigma}\tilde{\cal F}_{\mu\nu\rho\sigma}=0),\ee
where $\epsilon_{\mu\nu\rho\sigma}$ is the totally antisymmetric tensor with $\epsilon_{0123}$ equal to the metric determinant $g$, such that $g_{\mu\alpha}g_{\nu\beta}g_{\rho\gamma}g_{\sigma\delta}\epsilon^{\alpha\beta\gamma\delta}=\epsilon_{\mu\nu\rho\sigma}$.

All the ingredients introduced so far are sufficient to describe gravity only. However, we want to include all the other interactions (electroweak, strong, Yukawa interactions, etc.) so we also consider a generic number of real scalars (or pseudoscalars) $\phi$, gauge fields $A^I_\mu$ corresponding to an internal gauge group $G$  and fermions, which we represent here with Weyl spinors $\psi$. Note that massive vector fields 
can be obtained as usual through the Higgs or  St\"uckelberg mechanisms.

In general the distorsion tensor does not have special properties. However, in the presence of fermions one can show that it should be such that the covariant derivative of the metric vanishes, or, in other words, the theory should be metric compatible. 

As we will recover now, this has to do with the fact that 
in a generic curved spacetime fermion fields belong to the spinorial representation of a  local Lorentz group in the tangent space. Indeed in order to define them one introduces a basis $\{e_a\}$ in the tangent space such that 
\be \eta_{ab}=e_a^\mu e_b^\nu g_{\mu\nu}, \label{Ortho} \ee
where the ``tetrads" $e_a^\mu$ are defined by expanding each $e_a$ in the coordinate basis, $e_a=e_a^\mu \frac{\partial}{\partial x^\mu}$.
We can also define   $e^a_\mu\equiv \eta^{ab}g_{\mu\nu}e^\nu_b$, which can be considered as the components of some one-form fields $e^a$ in the one-form basis $\{dx^\mu\}$, namely $e^a= e^a_\mu dx^\mu$. Using~(\ref{Ortho}) one finds that these quantities satisfy  $e^a_\mu e^\mu_b=\delta^a_b$ and
\be g_{\mu\nu}=e^a_\mu e^b_\nu\eta_{ab}.\label{gfrome}\ee 
It follows that the inverse of the metric exists and is given by $g^{\mu\nu} = e_a^\mu e_b^\nu \eta^{ab}$, which implies $\eta^{ab}=e^a_\mu e^b_\nu g^{\mu\nu}$.
The $e^a$ (and analogously the $e_a$) are defined modulo local Lorentz transformations: if we redefine $e^{'a} = \Lambda^a_{~b}e^b$, where $\Lambda^a_{~b}$  
are the elements of a local Lorentz transformation, we obtain the same metric $g_{\mu\nu}=e^{'a}_\mu e^{'b}_\nu\eta_{ab}$. Let us consider now a vector ${\cal V}$, which we take to be $G$-invariant for simplicity, and expand it in the basis $\{e_a\}$, that is ${\cal V}= {\cal V}^a e_a$.  The components ${\cal V}^a$ belong to the vector representation of the local Lorentz group so  their covariant derivative 
\be {\cal D}_\mu {\cal V}^a = \partial_\mu {\cal V}^a+{\cal A}_{\mu~b}^{~\,a}{\cal V}^b\ee
 should  feature a connection  ${\cal A}_{\mu~b}^{~\,a}$ (known as the spin connection) whose values belong to the Lorentz algebra: defining ${\cal A}_{\mu}^{~\,ab}\equiv{\cal A}_{\mu~b}^{~\,c} \eta^{bc}$,  we can impose an antisymmetry with respect to the exchange of the flat indices $a, b$:
\be {\cal A}_{\mu}^{~\,ab}=-{\cal A}_{\mu}^{~\,ba}.\label{antConn}\ee 
The spin connection can be seen as the connection ${\cal A}_{\mu~\sigma}^{~\,\rho}$ rewritten using the tetrad basis and we can express one in terms of the other: this can be done by considering the covariant derivative ${\cal D} {\cal V}$ and writing the identities
\be {\cal D}_\mu {\cal V}^\rho ~ dx^\mu \otimes \frac{\partial}{\partial x^\rho}= {\cal D} {\cal V}={\cal D}_\mu {\cal V}^a ~ dx^\mu \otimes e_a = e_a^\rho {\cal D}_\mu {\cal V}^a ~ dx^\mu\otimes \frac{\partial}{\partial x^\rho}\ee 
which implies $ {\cal D}_\mu {\cal V}^\rho=e_a^\rho {\cal D}_\mu {\cal V}^a$. Using then~(\ref{CovDer}) and ${\cal V}^a=e^a_\lambda {\cal V}^\lambda$ one finds
\be {\cal A}_{\mu~b}^{~\,a}  = e^a_\nu{\cal A}_{\mu~\lambda}^{~\,\nu} e^\lambda_b- e^\lambda_b\partial_\mu e^a_\lambda.\label{omegaA} \ee 
From this result one can show
\be {\cal D}_\mu e_\nu^a \equiv \partial_\mu e_\nu^a -{\cal A}_{\mu~\nu}^{~\,\lambda} e^a_\lambda + {\cal A}_{\mu~b}^{~\,a} e^b_\nu = 0   \ee 
and, therefore, using~(\ref{gfrome}), the antisymmetry property~(\ref{antConn}) and the Leibniz rule we obtain ${\cal D}_\mu g_{\alpha\beta}=0$.

The above-mentioned local Lorentz group is precisely the one with respect to which fermions belong to the spinorial representation.   Therefore, we recover  the well-known result that in the presence of fermions, when this local Lorentz group is compulsory, the theory should be metric compatible. In the absence of fermions, on the other hand, one can have ${\cal D}_\mu g_{\alpha\beta}\neq 0$ and $T_{\mu\nu\rho}=0$, which is known as Palatini gravity.

   The gauge fields $A^I_\mu$, together with the connection  ${\cal A}_{\mu~\sigma}^{~\,\rho}$, allow us to define a covariant derivative with respect to both general coordinate transformations and elements of $G$, whose action on scalars and fermions reads
\be {\cal D}_\mu \phi = \partial_\mu \phi+ i \theta^I A^I_\mu \phi, \qquad {\cal D}_\mu\psi = \partial_\mu \psi + i t^IA^I_\mu\psi + \frac12 {\cal A}^{ab}_\mu \sigma_{ab}  \psi, \label{SF-tr}\ee 
where, recalling that we work with Weyl fermions,   $\sigma^{ab} \equiv \frac14 (\sigma^a\bar\sigma^b-\sigma^b\bar\sigma^a)$, also $\sigma^i\equiv -\bar\sigma^i$ (with $i=1,2,3$) are the Pauli matrices and $\sigma^0\equiv\bar\sigma^0\equiv1$ is the $2\times 2$ identity matrix.  
The gauge couplings are contained in the matrices $\theta^I$ and $t^I$, which are the generators of $G$ in the scalar and fermion representations, respectively.

We consider now the commutator of two covariant derivatives acting on a scalar field $\phi$:
\be [{\cal D}_\mu,{\cal D}_\nu]\phi = \left[iF_{\mu\nu}^I \theta^I-({\cal A}_{\mu~\nu}^{~\,\lambda}-{\cal A}_{\nu~\mu}^{~\,\lambda}){\cal D}_\lambda\right]\phi, \label{Commphi} \ee 
where
\be F_{\mu\nu}^I \equiv \partial_\mu A^I_\nu-\partial_\nu A^I_\mu-f^{KJI} A_\mu^KA_\nu^J\ee 
and the $f^{KJI}$ are the structure constants of $G$. Note that both $[{\cal D}_\mu,{\cal D}_\nu]\phi$ and  $({\cal A}_{\mu~\nu}^{~\,\lambda}-{\cal A}_{\nu~\mu}^{~\,\lambda}){\cal D}_\lambda\phi$ are  tensors and so, because of~(\ref{Commphi}), also the $F_{\mu\nu}^I$ must be tensors.  
 This shows that the expression of the field strength tensor of $A^I_\mu$ in the presence of a generic connection can be taken to be $F_{\mu\nu}^I$, namely the same as the one in flat space even if the connection is not the LC one.

\section{Theories with non-dynamical distorsion}\label{eq-theories}
Before going to the general characterization  of theories with non-dynamical distorsion  it is useful to recall the structure of metric theories.

Einstein's GR is the leading theory of this type in the low energy limit. Its action is the standard Einstein-Hilbert one 
\be S_{\rm EH} = \int d^4x \sqrt{-g}\left( \frac{\bp^2}{2} R - \Lambda \right), \label{EHt} \ee 
where $\bp$ is the reduced Planck mass and $\Lambda$ is the cosmological constant. We can also add  higher curvature terms to $S_{\rm EH} $,
\be \int d^4x \sqrt{-g}\left(a_2 R^2 +b_2 R_{\mu\nu}R^{\mu\nu} +\frac{a_3}{\bp^2} R^3 + ... +\frac{a_4}{\bp^4} (R_{\mu\nu}R^{\mu\nu})^2 +  ...  \right),\ee
where the $a_i$, $b_i$, etc. are freely adjustable dimensionless coefficients.

Furthermore, we can also add to the theory a generic matter sector with action $ S_{\rm matter} = \int d^4x \sqrt{-g}\mathscr{L}_{\rm matter}$, where $\mathscr{L}_{\rm matter}$
can contain (pseudo)scalar fields $\phi$, fermions $\psi$ and gauge fields $A_\mu^I$. Besides the standard renormalizable terms (which play the leading role in the low energy limit) $\mathscr{L}_{\rm matter}$ can also contain higher-order terms built  with $\phi$, $\psi$ and $A_\mu^I$ as well as $g_{\mu\nu}$. All these terms, of course, must be compatible with the given symmetries (general coordinate invariance, $G$ and possibly some global symmetries). For example, we can add to $\mathscr{L}_{\rm matter}$ terms of the form $(F_{\mu\nu}^IF^{I\mu\nu})^2$, $({D}_\mu \phi{D}^\mu \phi)^3$, $R {D}_\mu \phi{D}^\mu \phi$  etc. with, again, freely adjustable coefficients.  

Adopting the EFT point of view, the higher the dimensionality of the generic added term is the less relevance such term has at low energies.  Using the same reasoning, we do not add non-local terms too, because at sufficiently low energies any non-locality will be described by a series of local terms.

\subsection{General characterization}\label{Characterization1}

The purpose of this section is to identify the most general class  
 of (local effective field) theories of the type defined in Sec.~\ref{Ingredients} where the distorsion $C_{\mu~\sigma}^{~\,\rho}$ is not dynamical. These theories are those whose action can be brought into the form  
\be S_{\rm eq} = \int d^4x\sqrt{-g}\left({\cal F}_{\mu\nu\rho\sigma} {\cal T}^{\mu\nu\rho\sigma}(\Phi) + \Sigma(\Phi, {\cal D}\Phi, C)    \right), \label{Seq1}\ee 
where $\Phi$ represents the set of fields that are independent of $C_{\mu~\sigma}^{~\,\rho}$, namely
\be \Phi=\{g_{\mu\nu}, \phi, \psi, F_{\mu\nu}^I, ... \},\ee
the dots are curvatures and covariant derivatives of the previous fields constructed with the LC connection,
${\cal T}^{\mu\nu\rho\sigma}(\Phi)$ is a rank-four contravariant tensor that depends on $\Phi$ only (not on its derivatives) and  $\Sigma(\Phi, {\cal D}\Phi, C)$ is a quantity that depends on $\Phi$  
and  $C_{\mu~\sigma}^{~\,\rho}$ only. Note that ${\cal T}^{\mu\nu\rho\sigma}(\Phi)$ and $\Sigma(\Phi, {\cal D}\Phi, C)$ should also be invariant under gauge transformations of $G$  and possibly some global symmetries, if any.

The reason why the distorsion is not dynamical for theories of the form~(\ref{Seq1}) is because  the field equations of $C_{\mu~\sigma}^{~\,\rho}$ are purely algebraic in $C_{\mu~\sigma}^{~\,\rho}$. Indeed, the derivatives of the distorsion only appear in the first term proportional to ${\cal T}^{\mu\nu\rho\sigma}$ and they are first derivatives, so, after an integration by parts it is possible to make them act on ${\cal T}^{\mu\nu\rho\sigma}$ instead. Therefore, in principle, these equations can be solved exactly to find $C_{\mu~\sigma}^{~\,\rho}$ as a functional of $\Phi$. Once this is done, the theory with action $S_{\rm eq}$ can always be written as a metric theory, whose general form has been described at the beginning of this section~\ref{eq-theories}.

Note that the theory defined in~(\ref{Seq1}) is the most general one with non-dynamical distorsion.
The reason is that even setting $C_{\mu~\sigma}^{~\,\rho}=0$ one can recover the most general metric theory: this is because, as we have specified, the collective field $\Phi$ can also contain curvature tensors and covariant derivatives of $\phi, \psi, \overline \psi, F_{\mu\nu}^I$ constructed with the LC connection. If one allows now for a non-vanishing distorsion, one can anyhow express it in terms of $\Phi$ by using its field equations.

We can thus state that the theories with non-dynamical distorsion are those whose action is linear in the curvature ${\cal F}_{\mu\nu~~\sigma}^{~~~\rho}$ of the full connection ${\cal A}_{\mu~\sigma}^{~\,\rho}$ with the ``coefficients" of the linear terms, i.e. the tensor ${\cal T}^{\mu\nu\rho\sigma}(\Phi)$, being independent of the distorsion itself. 
This class of theories can be regarded as equivalent formulations of the most general metric theories with the given set of matter fields $\{\phi, \psi, A_\mu^I\}$ (for examples of equivalent formulations of specific metric theories see Refs.~\cite{BeltranJimenez:2017tkd,BeltranJimenez:2019esp,BeltranJimenez:2019odq,Dadhich:2012htv}).

\subsection{Theories with a falsely-dynamical distorsion}\label{fake}

It is important to note that some theories, despite not appearing of the form~(\ref{Seq1}), can be brought into that form with appropriate redefinitions. 

To illustrate this point let us consider as an example the case where the action is 
\be S = \int d^4x\sqrt{-g}\left({\cal F}_{\mu\nu\rho\sigma} {\cal T}^{\mu\nu\rho\sigma}(\Phi) + \Delta(\Phi, \alpha(\Phi){\cal R}+\beta(\Phi) {\cal R}')+ \Sigma(\Phi, {\cal D}\Phi)    \right), \label{Sf}\ee 
with
\be {\cal T}^{\mu\nu\rho\sigma}(\Phi) = \alpha(\Phi) g^{\mu\rho}g^{\nu\sigma} + \beta(\Phi)\frac{\epsilon^{\mu\nu\rho\sigma}}{\sqrt{-g}}\label{Tf} \ee
and $\Delta$ is a function of $\Phi$ and the specific combination $\alpha(\Phi){\cal R}+\beta(\Phi) {\cal R}'$ only, where $\alpha$ and $\beta$ are the {\it same} functions of $\Phi$ that appear in~(\ref{Tf}). Moreover, we take $\Phi$ independent of the curvature and covariant derivatives built with the LC connection and  $\Sigma$ independent of ${\cal D}g_{\mu\nu}$; also we take $\alpha$, $\beta$  and $\Delta$ independent of  the metric and impose the further constraint $1+\frac{\partial\Delta}{\partial z}(\Phi,z)>0$. In this specific case, using~(\ref{Tf}), the action   reads
\be S = \int d^4x\sqrt{-g}\left(\alpha(\Phi){\cal R}+\beta(\Phi) {\cal R}' + \Delta(\Phi, \alpha(\Phi){\cal R}+\beta(\Phi) {\cal R}')+ \Sigma(\Phi, {\cal D}\Phi)    \right). \label{Sf2}\ee 

Theories of this form actually belong to the class of~(\ref{Seq1}) and, therefore, feature a non-dynamical distorsion.  In order to show that we introduce an auxiliary field $z$ that allows us to write $S$ in the form
\bea S &=& \int d^4x\sqrt{-g}\left(\alpha(\Phi)(1+\frac{\partial\Delta}{\partial z}(\Phi,z)){\cal R}+\beta(\Phi)(1+\frac{\partial\Delta}{\partial z}(\Phi,z)){\cal R'} \right. 
\nonumber \\ && \left.+ \Delta(\Phi,z)-z\frac{\partial\Delta}{\partial z}(\Phi,z)+\Sigma(\Phi,{\cal D}\Phi)\right). \label{Sf3}
\eea 
The action above is equivalent to the one in~(\ref{Sf2}) because  of the following argument. First note that we can impose the condition $\frac{\partial^2\Delta}{\partial z^2}\neq 0$ without loss of generality given that around any point where $\frac{\partial^2\Delta}{\partial z^2} = 0 $ we can write 
\be \Delta(\Phi, \alpha(\Phi){\cal R}+\beta(\Phi) {\cal R}') \simeq \Delta_0(\Phi) +\Delta_1(\Phi) ( \alpha(\Phi){\cal R}+\beta(\Phi) {\cal R}') \ee
and the functions of $\Phi$ that we called here $\Delta_0$ and $\Delta_1$ can be absorbed in an appropriate redefinition of $\alpha$, $\beta$ and $\Sigma$.
Now, by using the field equation of $z$ computed using the action in~(\ref{Sf3}), we find
\be \frac{\partial^2\Delta}{\partial z^2}(\Phi,z) (\alpha(\Phi){\cal R}+\beta(\Phi) {\cal R}' - z) = 0, \ee 
which implies, using $\frac{\partial^2\Delta}{\partial z^2}\neq 0$, that $z=\alpha(\Phi){\cal R}+\beta(\Phi) {\cal R}'$. By inserting this result  in~(\ref{Sf3}) one recovers exactly~(\ref{Sf2}). 

The reason why these theories can be brought into the form~(\ref{Seq1}) is because we can absorb the dependence on $z$ in front of both ${\cal R}$ and ${\cal R'}$ in~(\ref{Sf3}) through the metric rescaling
\be g_{\mu\nu}\to \Omega^2 g_{\mu\nu}, \label{rescalingg}\ee
 where $\Omega^2$ depends only algebraically on $z$:
 \be \Omega^2(\Phi,z) = \frac{1}{1+\frac{\partial\Delta}{\partial z}(\Phi,z)}\label{rescalingg2}\ee 
 (here is where we use $1+\frac{\partial\Delta}{\partial z}(\Phi,z)>0$).
After this metric rescaling the spacetime derivatives of $z$ do not appear because we do not change at the same time\footnote{The spin connection in~(\ref{omegaA}) transforms as ${\cal A}^{ab}_\mu\to{\cal A}^{ab}_\mu-\eta^{ab}\partial_\mu\ln\Omega$ under the metric rescaling~(\ref{rescalingg}) (see~\cite{Chakrabarty:2018ybk} for a related study) so that also the covariant derivative of fermions in~(\ref{SF-tr}) is invariant.}
${\cal A}_{\mu~\sigma}^{~\,\rho}$, $\phi$, $\psi$ and $A_\mu^I$ and, as specified, we take $\Phi$ independent of the curvature and covariant derivatives built with the LC connection and  $\Sigma$ independent of ${\cal D}g_{\mu\nu}$. Therefore, we can easily integrate out $z$ and express it in terms of the other fields $\Phi$. So in this case $z$ is not dynamical and there are no other degrees of freedom besides the metric and the matter fields $\{\phi, \psi, A_\mu^I\}$.

\subsection{$f({\cal R})$ theories}

A particular form of Eq.~(\ref{Sf2}) is 
\be S = \int d^4x\sqrt{-g} f({\cal R}),\ee
 where $f$ is a function with $\frac{\partial f}{\partial {\cal R}} > 0$ and $\frac{\partial^2 f}{\partial {\cal R}^2} \neq 0$. Therefore, we obtain that also  $f({\cal R})$ metric-affine theories\footnote{For a specific treatment of $f({\cal R})$ see e.g. Refs.~\cite{Sotiriou:2008rp,DeFelice:2010aj,Rodrigues:2011zi}.} do not feature a dynamical distorsion.  

Also, as a consequence of the calculations we have performed in Sec.~\ref{fake}, the $f({\cal R})$ metric-affine theories can actually be recast in the GR form~(\ref{EHt}).  Indeed, by defining the function $\Delta$ through
\be \alpha {\cal R} +\Delta(\alpha {\cal R}) \equiv f({\cal R}), \ee
where $\alpha$ is an arbitrary positive constant, we obtain (after the metric rescaling in~(\ref{rescalingg}) and~(\ref{rescalingg2}))
\be S = \int d^4x\sqrt{-g}\left(\alpha {\cal R} +\alpha^2\frac{f(\tilde z)-\tilde z \frac{\partial f}{\partial \tilde z}(\tilde z)}{\frac{\partial f}{\partial \tilde z}(\tilde z)^2} \right), \ee
where $\tilde z\equiv z/\alpha$; the field $\tilde z$ is clearly non dynamical and in principle we can solve its field equation and plug the solution into the action to obtain
\be S = \int d^4x\sqrt{-g}\left(\alpha {\cal R} - \Lambda\right), \ee
where 
\be \Lambda =\alpha^2\frac{\tilde z_0 \frac{\partial f}{\partial \tilde z_0}(\tilde z_0)-f(\tilde z_0)}{\frac{\partial f}{\partial \tilde z_0}(\tilde z_0)^2} \ee 
and $\tilde z_0$ is a solution of the $\tilde z$ field equation. After that, using the techniques in Appendix~\ref{appendixInt} (see also Ref.~\cite{Dadhich:2012htv}), we can solve the field equations of the distorsion and insert the solution into the action to obtain precisely~(\ref{EHt}), with the identification $\alpha =\bp^2/2$.

 This means, among other things, that   $f({\cal R})$ metric-affine theories do not have any other gravitational degrees of freedom besides the ordinary graviton (see also~\cite{Antoniadis:2018ywb} for a previous related discussion, and~\cite{Edery:2019txq} for the particular  case $f({\cal R}) \propto {\cal R}^2$). Instead, in $f(R)$ metric theories the gravitational spectrum features, in additional to the ordinary graviton, a dynamical scalar field: technically this happens because it is not possible to rescale the metric as in~(\ref{rescalingg}) without changing the connection in the metric case (where the connection is the LC one).

\section{Theories with dynamical distorsion}\label{Dyn-Dist}

\subsection{General characterization}

The general form~(\ref{Seq1}) of theories with non-dynamical distorsion is useful, among other things, because it helps us in identifying  the class of theories with a dynamical distorsion: they are those that can never be brought into the form~(\ref{Seq1}). Indeed, in this case kinetic terms for some components of the distorsion necessarily appear. In general there can be other components of the distorsion that remain non dynamical: we say that the distorsion is dynamical when at least some components of this tensor feature kinetic terms.

In the following we provide some examples of (local effective field) theories that cannot be brought into the form~(\ref{Seq1}) and, in simple cases, compute explicitly the kinetic terms for the dynamical  components of the distorsion.

\subsection{Examples: Dynamical (pseudo)scalarons}\label{(pseudo)scalaron}

As we have seen, the theories with non-dynamical distorsion are those whose action can be brought into a form that is linear in the curvature  of the full connection  with the ``coefficients" of the linear terms  being independent of the distorsion itself. Therefore, generically, we can have a dynamical distorsion by adding terms that are non linear in the curvature.
So the first examples of metric-affine theories with dynamical distorsion that we consider have actions of the form

\be S = \int d^4x\sqrt{-g}\left({\cal F}_{\mu\nu\rho\sigma} {\cal T}^{\mu\nu\rho\sigma}(\Phi) + \Delta(\Phi, {\cal R}, {\cal R}')+ \Sigma(\Phi, {\cal D}\Phi, C)    \right), \label{Seq}\ee 
where $\Phi$, ${\cal T}^{\mu\nu\rho\sigma}(\Phi)$ and  $\Sigma(\Phi, {\cal D}\Phi, C)$ have been defined in Sec.~(\ref{Characterization1}) and $\Delta$ 
 is a function of $\Phi, {\cal R}$ and ${\cal R}'$ only. Note that  $\Delta(\Phi, {\cal R}, {\cal R}')$  should also be invariant under gauge transformations of $G$ and the global symmetries, if any. The function $\Delta$ can introduce the non linearity in the curvature that is crucial to have a dynamical distorsion. Indeed, barring specific choices of the action, such as those described in Sec.~\ref{fake}, one has dynamical (pseudo)scalar degrees of freedom coming from the distorsion in this case, as we now show.

Let us start with the case in which $\Delta$ does not depend on ${\cal R}'$, but can have a generic dependence on $\Phi$ and ${\cal R}$. This case can be treated by introducing one auxiliary scalar field $\zeta$. The action $S$ can be equivalently written  as follows
\be S = \int d^4x\sqrt{-g}\left({\cal F}_{\mu\nu\rho\sigma} {\cal T}^{\mu\nu\rho\sigma}(\Phi) + \Delta(\Phi, \zeta)+\frac{\partial\Delta}{\partial\zeta}(\Phi, \zeta)({\cal R}-\zeta)+ \Sigma(\Phi, {\cal D}\Phi, C)    \right). \label{Seq2}\ee 
To show this we observe that the field equation of $\zeta$ is 
\be 
({\cal R}-\zeta)\frac{\partial^2\Delta}{\partial\zeta^2}=0. \label{EOMzeta}\ee
and that we can require without loss of generality  $\frac{\partial^2\Delta}{\partial\zeta^2}\neq 0$. Indeed, around any point with $\frac{\partial^2\Delta}{\partial\zeta^2}= 0$
 we can have at most a linear dependence of $\Delta$ on 
 ${\cal R}$, and we can, therefore, absorb $\Delta$ in a redefinition of ${\cal T}^{\mu\nu\rho\sigma}$  and $\Sigma$ and go back to the case of non-dynamical distorsion of Sec.~\ref{eq-theories}. From~(\ref{EOMzeta}) it follows that the field equations fix $\zeta={\cal R}$ and~(\ref{Seq2}) reduces to~(\ref{Seq}). We can now write 
 \be S = \int d^4x\sqrt{-g}\left({\cal  F}_{\mu\nu\rho\sigma} {\cal \bar T}^{\mu\nu\rho\sigma}(\Phi,\zeta) + \bar \Sigma(\Phi, \zeta, {\cal D}\Phi,C)    \right), \label{Seq3}\ee 
 where
 \bea {\cal \bar T}^{\mu\nu\rho\sigma}(\Phi,\zeta) &\equiv& {\cal T}^{\mu\nu\rho\sigma}(\Phi) + g^{\mu\rho}g^{\nu\sigma}\frac{\partial\Delta}{\partial \zeta}(\Phi,\zeta), \\
 \bar \Sigma(\Phi,\zeta,{\cal D}\Phi, C) &\equiv&  \Sigma(\Phi,{\cal D}\Phi, C) + \Delta(\Phi,\zeta)-\zeta\frac{\partial\Delta}{\partial\zeta}(\Phi,\zeta).\eea
 Therefore, we have come back to the previously studied case $\Delta=0$, but with an extra scalar $\zeta$ in addition to the $\phi$ fields we started with. 
 In deriving the algebraic equations of $C_{\mu~\sigma}^{~\,\rho}$, derivatives of $\zeta$ generically appear  when we integrate by parts the terms coming from ${\cal  F}_{\mu\nu\rho\sigma} {\cal \bar T}^{\mu\nu\rho\sigma}(\Phi,\zeta)$ that contain one derivative of the variation of $C_{\mu~\sigma}^{~\,\rho}$, see Eq.~(\ref{FRC}). This fact can produce a kinetic term for $\zeta$, barring specific choices of the action. An example of such specific choices is when ${\cal  T}^{\mu\nu\rho\sigma}\propto g^{\mu\rho}g^{\nu\sigma}$ as we have seen in Sec.~\ref{fake}.

Whether this new dynamical scalar  $\zeta$ is a manifestation of the dynamics of the distorsion is not clear. This is because ${\cal R}$, which is equal to $\zeta$ by using the field equations, does not vanish when the distorsion is zero (see Eq.~(\ref{RRC})) and so a dynamical $\zeta$ could also correspond  just  to an extra dynamical  scalar  from the metric. 
 
 Since this section is devoted to theories with a dynamical distorsion we then 
   consider the case where $\Delta$ depends on both $\cal R$ and $\cal R'$, but for now only through a linear combination \be \rho\equiv a(\Phi){\cal R}+b(\Phi)\cal R'.\label{rhodef}\ee
   Note that this situation is a generalization of the theories with a falsely-dynamical distorsion that we have analyzed in Sec.~\ref{fake}, where $a=\alpha$, $b=\beta$ and ${\cal T}^{\mu\nu\rho\sigma}$ was chosen to be of the specific type~(\ref{Tf}).  From the technical point of view this case can be treated similarly, but, as we will see soon, generically there is one more dynamical scalar here. Again we introduce an auxiliary field $z$ and we can show that $S$ can be equivalently written as 
 \be S = \int d^4x\sqrt{-g}\left({\cal F}_{\mu\nu\rho\sigma} {\cal T}^{\mu\nu\rho\sigma}(\Phi) + \Delta(\Phi, z)+\frac{\partial\Delta}{\partial z}(\Phi, z)(\rho-z)+ \Sigma(\Phi, {\cal D}\Phi, C)    \right) \label{Seq4}\ee 
 if the non-restrictive condition $\frac{\partial^2\Delta}{\partial z^2}\neq 0$ is imposed.  At this point we can again write $S$ as in~(\ref{Seq3}) 
 but with different redefined tensors:
 \bea {\cal \bar T}^{\mu\nu\rho\sigma}(\Phi,z) &\equiv& {\cal T}^{\mu\nu\rho\sigma}(\Phi) + \left(g^{\mu\rho}g^{\nu\sigma}a(\Phi)+ \frac{\epsilon^{\mu\nu\rho\sigma}}{\sqrt{-g}}b(\Phi)\right)\frac{\partial\Delta}{\partial  z}(\Phi,z), \\
 \bar \Sigma(\Phi,z,{\cal D}\Phi, C) &\equiv&  \Sigma(\Phi,{\cal D}\Phi, C) + \Delta(\Phi,z)-z\frac{\partial\Delta}{\partial z}(\Phi,z).\eea
So also here we have come back to the previously studied case $\Delta=0$, but with a new scalar $z$. Again, barring specific choices of the action (e.g. the ones of Sec.~\ref{fake}), the kinetic term of $z$ generically emerge when we solve for the distorsion because of the term ${\cal  F}_{\mu\nu\rho\sigma} {\cal \bar T}^{\mu\nu\rho\sigma}(\Phi,\zeta)$, which contains one derivative of $C_{\mu~\sigma}^{~\,\rho}$. When the kinetic term appears the field $z$ shows its dynamical nature, but again it is not clear whether this dynamics comes from the distorsion or from the metric because, using the field equations, $z=\rho$ and Eqs.~(\ref{rhodef}),~(\ref{RRC}) and~(\ref{RpRC}) tell us that a part of this dynamical field is sourced by the metric and a part is sourced by the distorsion. 

A class of theories where the distorsion is certainly dynamical can be found by considering the generic case where the dependence of $\Delta$ on $\cal R$ and $\cal R'$ is arbitrary. 
This case can be treated by introducing an auxiliary scalar field $\zeta$ and an auxiliary pseudoscalar field $\zeta'$. The action  can be equivalently written  as follows
\bea S &=& \int d^4x\sqrt{-g}\left({\cal F}_{\mu\nu\rho\sigma} {\cal T}^{\mu\nu\rho\sigma}(\Phi) + \Delta(\Phi, \zeta,\zeta')\right.\nonumber \\&&\left.+\frac{\partial\Delta}{\partial\zeta}(\Phi,\zeta,\zeta')({\cal R}-\zeta)+\frac{\partial\Delta}{\partial\zeta'}(\Phi,\zeta,\zeta')({\cal R'}-\zeta')+ \Sigma(\Phi, {\cal D}\Phi, C)    \right). \label{Seq5}\eea 
To show this we observe that the field equations of $\zeta$ and $\zeta'$ are, respectively, 
\bea  ({\cal R}-\zeta)\frac{\partial^2\Delta}{\partial\zeta^2}+({\cal R'}-\zeta')\frac{\partial^2\Delta}{\partial\zeta'\partial\zeta}&=&0 \\ 
({\cal R}-\zeta)\frac{\partial^2\Delta}{\partial\zeta'\partial\zeta}+({\cal R'}-\zeta')\frac{\partial^2\Delta}{\partial\zeta'^2}&=&0.\eea
Therefore, when the Hessian matrix of $\Delta$ (with respect to the variables $\zeta$ and $\zeta'$) is not singular these field equations imply ${\cal R}=\zeta$ and ${\cal R'}=\zeta'$ and~(\ref{Seq5}) is equivalent to~(\ref{Seq}). 
 We can always require that the Hessian matrix of $\Delta$ is not singular without loss of generality because around any point where this matrix is singular $\Delta$ depends at most linearly on a linear combination of ${\cal R}$ and ${\cal R'}$ (with a coefficient independent of the other linearly independent combination) and we can go  back to the previously analysed cases with a redefinition of ${\cal T}^{\mu\nu\rho\sigma}$. Now we can again write the action as in~(\ref{Seq3}), 
 but  with the following redefined tensors that this time depend on both $\zeta$ and $\zeta'$:
 \bea {\cal \bar T}^{\mu\nu\rho\sigma}(\Phi,\zeta,\zeta') &\equiv& {\cal T}^{\mu\nu\rho\sigma}(\Phi) + g^{\mu\rho}g^{\nu\sigma} \frac{\partial\Delta}{\partial \zeta}(\Phi,\zeta,\zeta')+ \frac{\epsilon^{\mu\nu\rho\sigma}}{\sqrt{-g}} \frac{\partial\Delta}{\partial \zeta'}(\Phi,\zeta,\zeta'), \\
 \bar \Sigma(\Phi,\zeta,\zeta',{\cal D}\Phi,C) &\equiv&  \Sigma(\Phi,{\cal D}\Phi,C) + \Delta(\Phi,\zeta,\zeta')-\zeta\frac{\partial\Delta}{\partial \zeta}(\Phi,\zeta,\zeta')-\zeta'\frac{\partial\Delta}{\partial\zeta'}(\Phi,\zeta,\zeta').\eea
So, again, we have come back to the previously studied case $\Delta=0$, but with the new scalars $\zeta$ and $\zeta'$ and when we derive the algebraic equations of $C_{\mu~\sigma}^{~\,\rho}$ derivatives of both $\zeta$ and $\zeta'$ appear in integrating by parts the terms coming from ${\cal  F}_{\mu\nu\rho\sigma} {\cal \bar T}^{\mu\nu\rho\sigma}(\Phi,\zeta,\zeta')$. So, generically, both $\zeta$ and $\zeta'$ can be dynamical, barring specific choices of the action\footnote{Note that whenever $\zeta$ and $\zeta'$ are non dynamical they can be integrated out and this leads to an equivalent metric theory, which could have been obtained even without the $\Delta$ term.  }.

\vspace{0.2cm}

The fields $\zeta$ and $\zeta'$  have a purely geometrical origin. We refer to them as the scalaron and the pseudoscalaron, respectively. The pseudoscalaron is particularly interesting for our purposes because it corresponds to a  degree of freedom coming essentially from the distorsion: using the field equations $\zeta'={\cal R'}$ and, according to Eq.~(\ref{RpRC}), ${\cal R'}$ can be non zero only if the distorsion is not zero. As discussed above $\zeta$ and $\zeta'$  are generically dynamical, but computing explicitly the corresponding  kinetic and interaction terms is of course very difficult and not very illuminating in the most general case of~(\ref{Seq}). Therefore, from now on to study the (pseudo)scalaron we focus on a less general class of theories. We take an action of the form 
\be S = \int d^4x\sqrt{-g}\left(\alpha(\Phi){\cal R}+\beta(\Phi){\cal R'} + \Delta(\Phi,{\cal R}, {\cal R'})+\Sigma(\Phi,{\cal D}\Phi)\right), \label{Seq-DN}\ee
where $\alpha$ and $\beta$ are functions of $\Phi$. Also, for simplicity, we take $\Phi$ independent of the curvature and covariant derivatives built with the LC connection and $\Sigma$  independent of ${\cal D}g_{\mu\nu}$. This is clearly a particular case of~(\ref{Seq}).

\subsubsection{Dynamical pseudoscalaron $\zeta'$}\label{Dynamicalzp}

 Let us now provide explicit examples of the most interesting case where $\zeta'$ is dynamical and explicitly compute its kinetic and potential terms. 
 
 To simplify the calculation of the kinetic and potential terms of $\zeta'$ here we also assume  that $\Delta$ is independent of ${\cal R}$  and   that there are no matter fields $\{\phi, \psi, A_\mu^I\}$, so that we can drop $\Sigma$ and  write
\bea S &=& \int d^4x\sqrt{-g}\left( \alpha{\cal R}+\beta {\cal R'} + \Delta({\cal R'})\right) \nonumber\\
&=& \int d^4x\sqrt{-g}\left[ \alpha{\cal R}+\left(\beta+\frac{\partial \Delta}{\partial\zeta'}(\zeta')\right) {\cal R'} + \Delta(\zeta') -\zeta'\frac{\partial \Delta}{\partial\zeta'}(\zeta')\right]  \label{Szetap}\eea
having required, again without loss of generality, $\frac{\partial^2\Delta}{\partial \zeta'^2}\neq 0$. The quantities $\alpha$ and $\beta$ are real parameters here; we will shortly identify $\alpha = M^2_P/2$ so we also have to assume $\alpha>0$; the ratio $\bp^2/(4\beta)$ is also known as the  Barbero-Immirzi parameter.
In this case, unlike those discussed in Sec.~\ref{fake}, it is not possible to have the quantities in front of both ${\cal R}$ and ${\cal R'}$ constant after a metric rescaling and $\zeta'$ becomes dynamical. Indeed, by using~(\ref{RRC}) and~(\ref{RpRC}) and integrating out $C_{\mu~\sigma}^{~\,\rho}$ leads to (see Appendix~\ref{appendixInt})    
\be S = \int d^4x\sqrt{-g}\left[\alpha R-K(\zeta')\frac{(\partial \zeta')^2}{2} -U(\zeta') \right], \label{SeqzpD}\ee
where we have defined \be K(\zeta') = \frac{24\bp^2}{1+16 B^2(\zeta')}\left(\frac{\partial B}{\partial \zeta'}\right)^2, \qquad B(\zeta')= \frac{\beta+\frac{\partial \Delta}{\partial \zeta'}(\zeta')}{M_P^2},\qquad U(\zeta') = \zeta'\frac{\partial \Delta}{\partial\zeta'}(\zeta')-\Delta(\zeta').\label{KBUzetap}\ee 
This is a standard Einstein-Hilbert action plus a kinetic and potential terms for an ordinary matter field. So we have to identify 
\be \alpha =\frac{\bp^2}{2}.\ee
Note that $B$ has to depend non-trivially on $\zeta'$ because of $\frac{\partial^2\Delta}{\partial \zeta'^2}\neq 0$.
 The second term in~(\ref{SeqzpD}) is a kinetic term of $\zeta'$, which is therefore dynamical. Note that $K(\zeta')$ is  always positive, so $\zeta'$ is never a ghost. 
We can render the kinetic term of this dynamical scalar canonical through the field redefinition
 \be \omega(\zeta') = \int_0^{\zeta'} dx \sqrt{K(x)}. \label{omegazeta}\ee 
Indeed, calling $\zeta'(\omega)$  the inverse function, which is uniquely defined  because $\frac{d\omega}{d\zeta'} = \sqrt{K} > 0$, and inserting in~(\ref{SeqzpD}) one obtains
  \be S = \int d^4x\sqrt{-g}\left[\frac{M_P^2}{2} R-\frac{(\partial \omega)^2}{2} -U(\zeta'(\omega)) \right].\label{SeqCan}\ee

 The provided examples where $\zeta'$ is dynamical are very interesting because, as mentioned above, ${\cal R'}$ is non-vanishing only when $C_{\mu~\sigma}^{~\,\rho}$ is present; so in these cases the distorsion has a scalar dynamical component.
 Given the relevance of this case we look for a general expression for the mass of $\zeta'$ (defined as the mass of the fluctuations of $\zeta'$ around a Lorentz invariant solution). First note that a Lorentz invariant stationary point of $S$ with respect to $\zeta'$  has to be a stationary point of $\Delta-\zeta'\frac{\partial \Delta}{\partial\zeta'}$, that is a solution of
\be \zeta'\frac{\partial^2 \Delta}{\partial\zeta'^2} = 0. \ee
But $\frac{\partial^2\Delta}{\partial \zeta'^2}\neq 0$ so the only Lorentz invariant stationary point is $\zeta' =0$. This can be understood observing that the field equations fix ${\cal R'}=\zeta'$ and   Lorentz invariance requires $C_{\mu~\sigma}^{~\,\rho}=0$, which implies ${\cal R'}=0$ according to Eq.~(\ref{RpRC}). 
 Note that Lorentz invariance also requires $\Delta-\zeta'\frac{\partial\Delta}{\partial\zeta'}=0$ and so, using $\zeta'=0$, one obtains $\Delta(0)=0$. To compute the mass of $\zeta'$ around $\zeta'=0$ we can focus on the part of the Lagrangian in~(\ref{SeqzpD}) that is quadratic in $\zeta'$,
\be -\frac{24\bp^2}{(1+16 B^2(0))}\left(\frac{\partial B}{\partial\zeta'}(0)\right)^2\frac{(\partial \zeta')^2}{2}  -\frac12 \frac{\partial^2 \Delta}{\partial\zeta'^2}(0) \zeta'^2. \ee
So the squared mass of $\zeta'$ is
\be m^2_{\zeta'} = \frac{(1+16 B^2(0))\frac{\partial^2 \Delta}{\partial\zeta'^2}(0)}{24\bp^2\left(\frac{\partial B}{\partial\zeta'}(0)\right)^2}. \label{mzp2}\ee 
We observe that $m^2_{\zeta'} \neq 0$ as a consequence of $\frac{\partial^2\Delta}{\partial \zeta'^2}\neq 0$, which also implies $\frac{\partial B}{\partial\zeta'}\neq 0$, so that the denominator in~(\ref{mzp2}) never vanishes. The requirement that $\zeta'$ is not a tachyon leads to the condition $\frac{\partial^2 \Delta}{\partial\zeta'^2}(0)> 0$. 

\begin{figure}[t]
\begin{center}
 \includegraphics[scale=0.37]{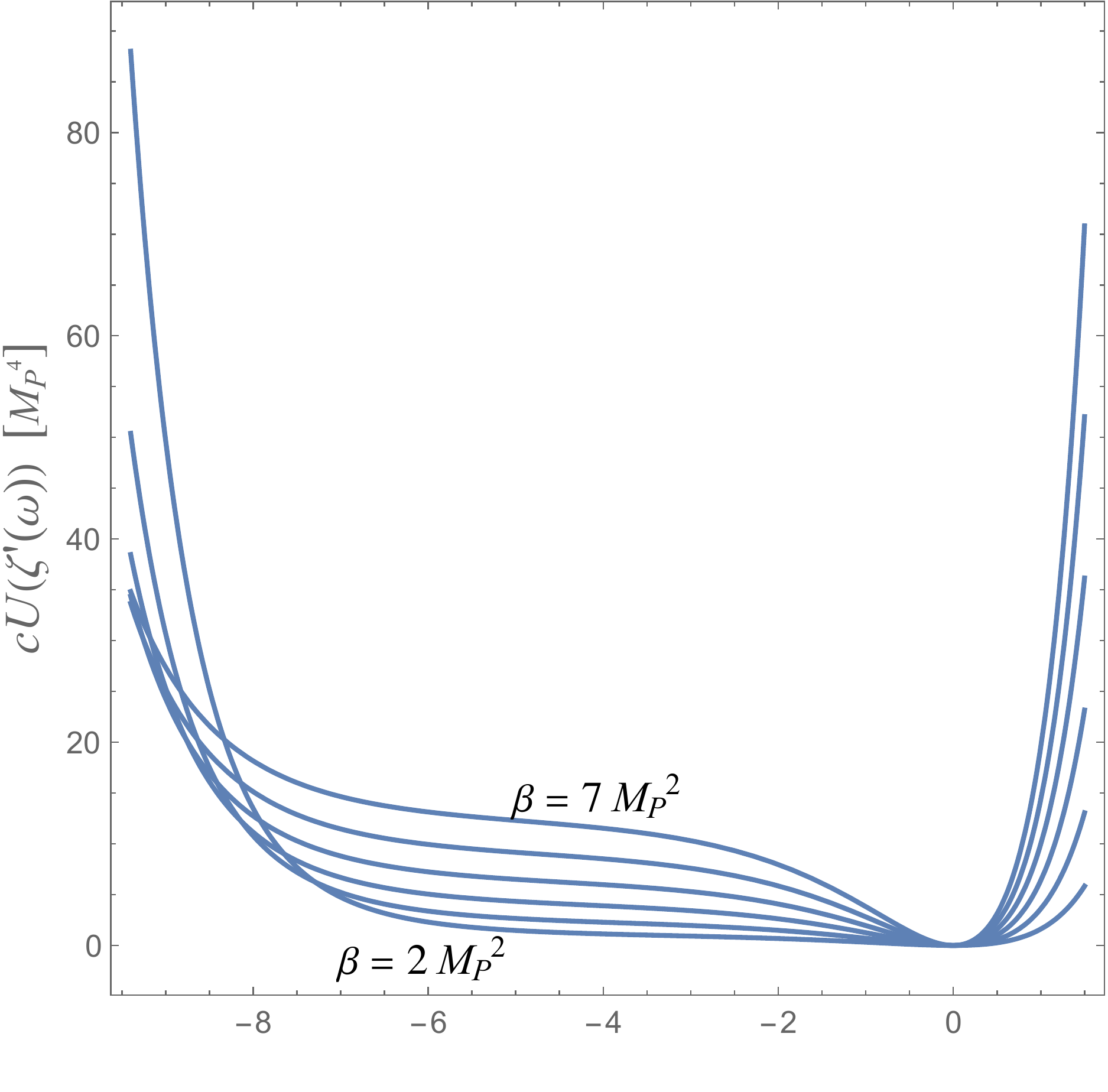}  
  \hspace{1cm}   \includegraphics[scale=0.37]{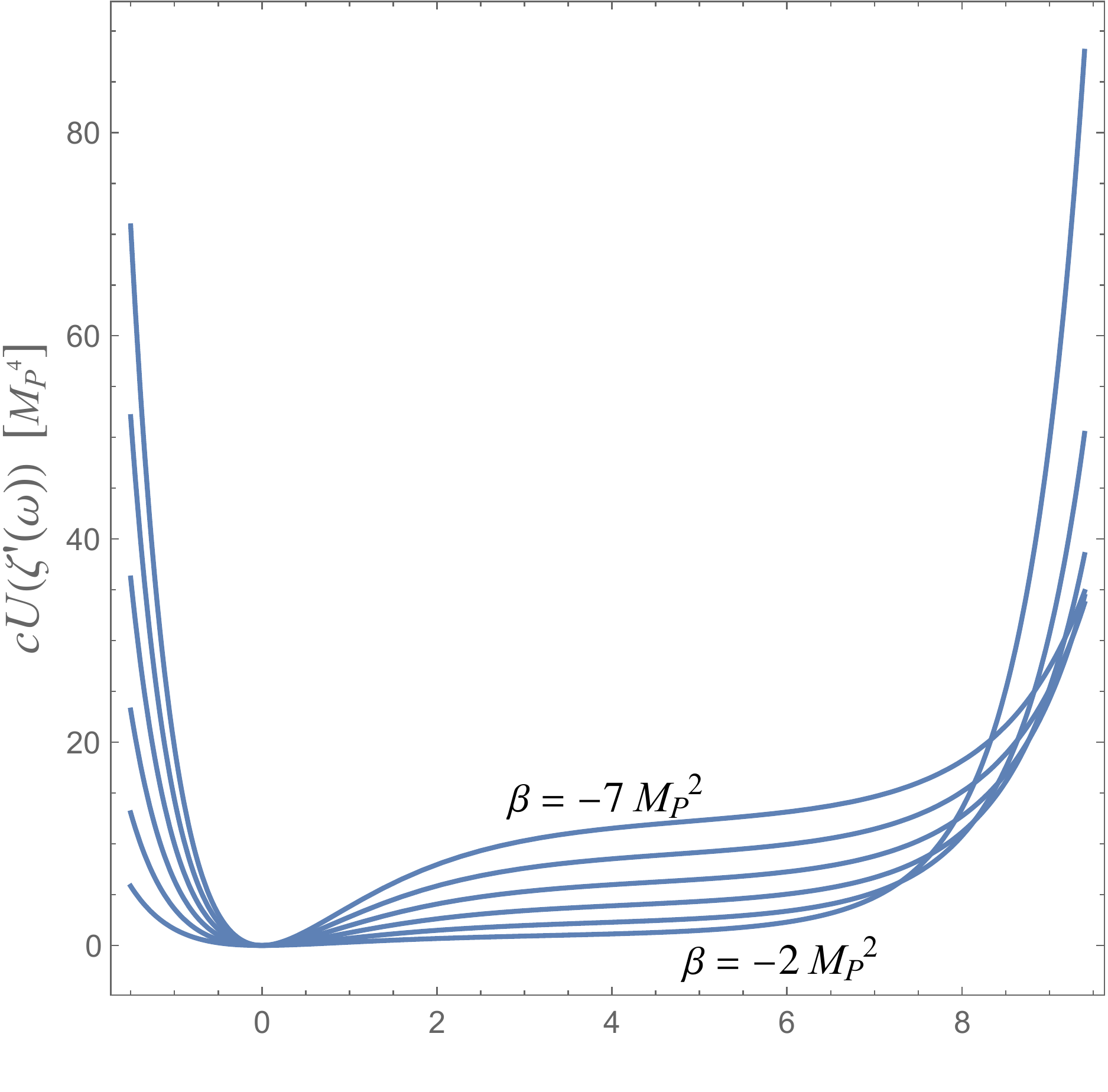} \\
   \hspace{-0.5cm}  \includegraphics[scale=0.37]{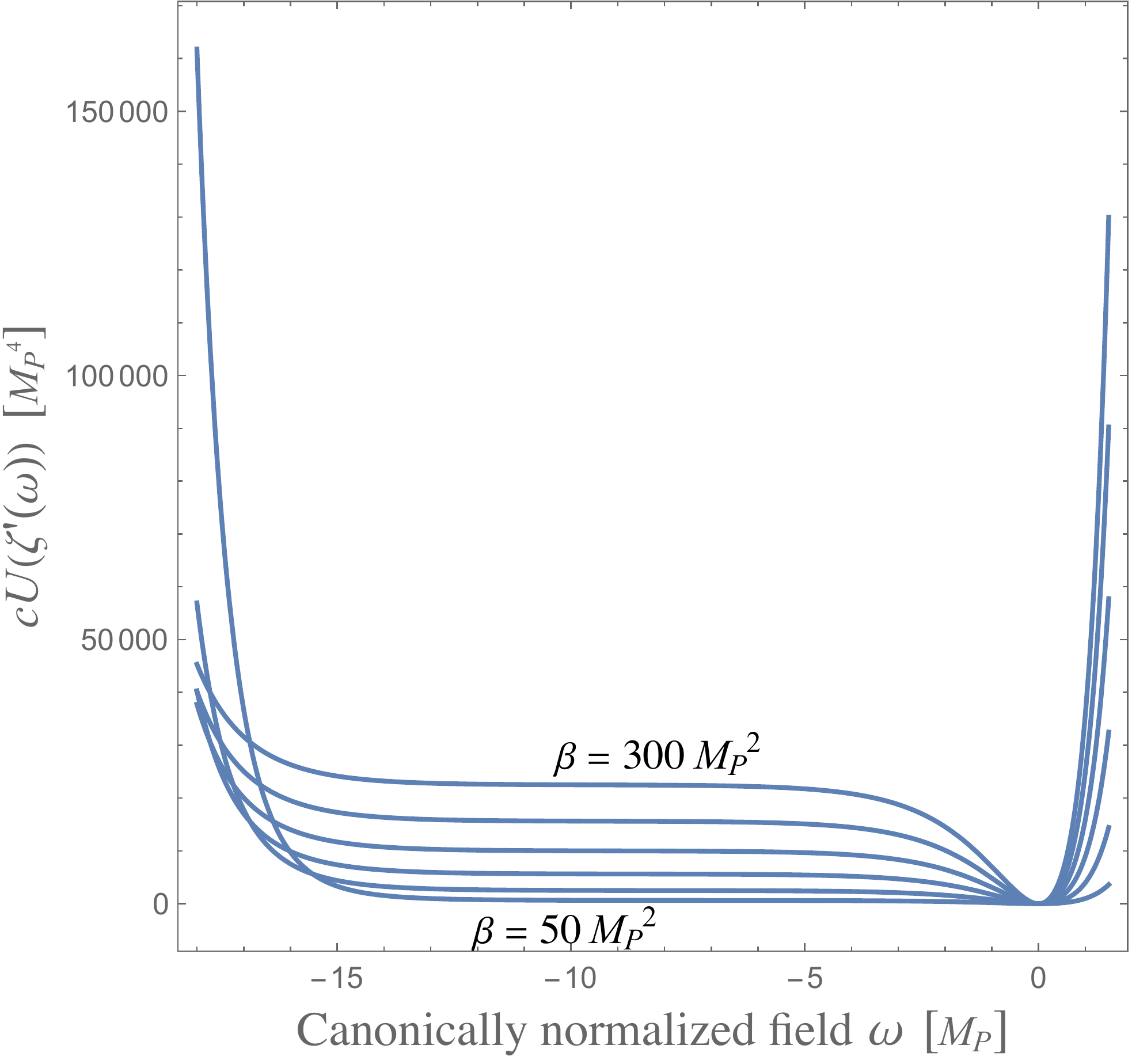}  
  \hspace{0.5cm}   \includegraphics[scale=0.37]{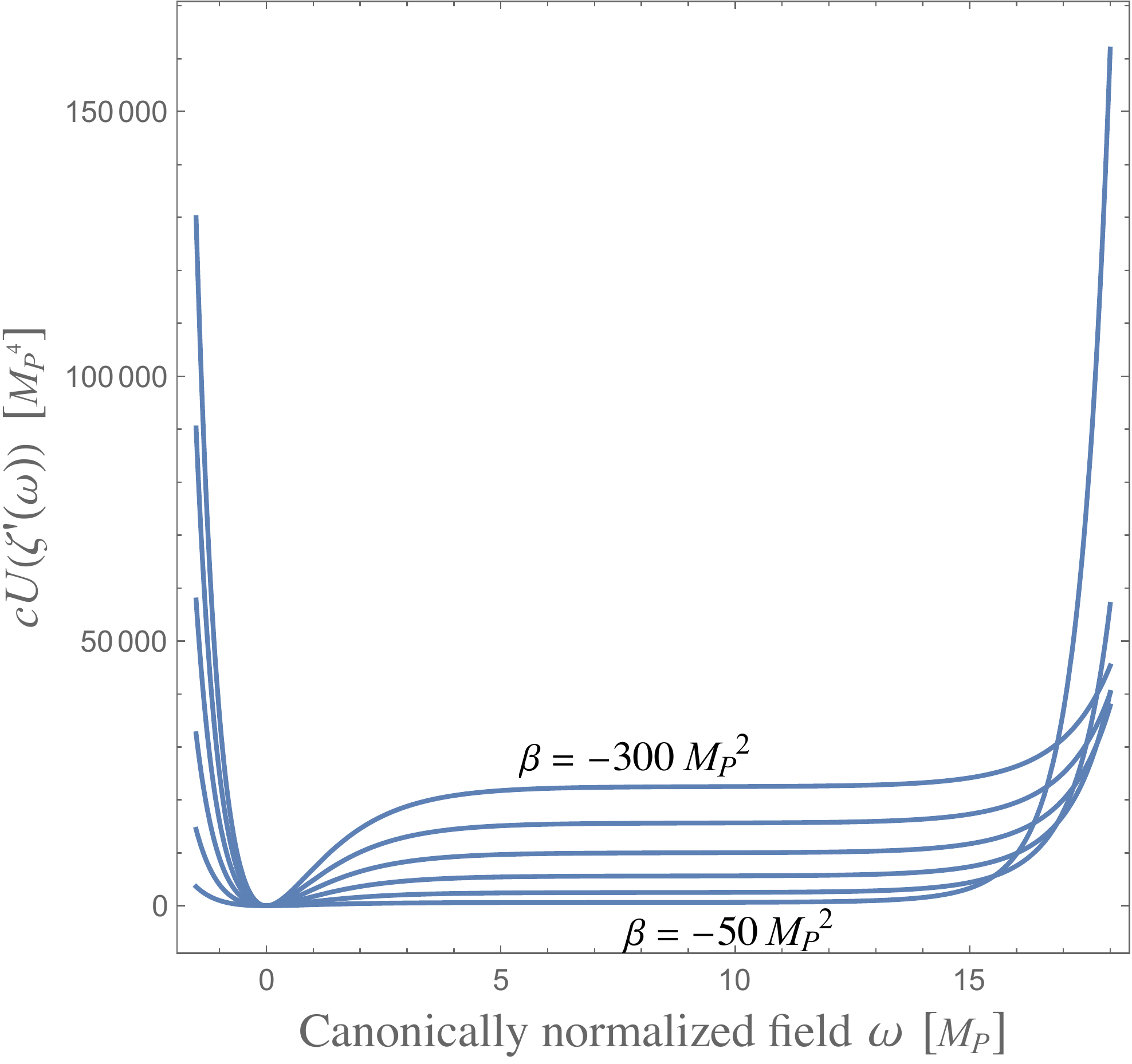} 
 	 
 \end{center}
   
   \caption{\em Potential of the canonically normalized pseudoscalaron (for  $\Delta({\cal R'}) = c {\cal R'}^2$) multiplied by $c$.   {\bf Left plots:} positive values of $\beta$. {\bf Right plots:} negative values of $\beta$.}\vspace{0.1cm}
\label{potzetap}
\end{figure}

The potential $U(\zeta'(\omega))$ can only   be explicitly computed once the function $\Delta$ is specified. Let us consider, for example\footnote{The ${\cal R'}^2$ term has appeared in different models in the literature, see e.g.~\cite{Hecht:1996np,BeltranJimenez:2019hrm,Dombriz:2021bnl}.
},  $\Delta({\cal R'}) = c {\cal R'}^2$ , where $c$  is a  positive constant (so that $\zeta'$ is not a tachyon). In this case we obtain 
\be  B(\zeta') = \frac{\beta+2c\zeta'}{\bp^2},\quad \frac{\partial B}{\partial\zeta'} =\frac{2c}{\bp^2}, \quad K(\zeta') =\frac{96 c^2}{\bp^2 \left[1+\frac{16 (2c \zeta'+\beta)^2}{\bp^4}\right]}, \quad  U(\zeta') =c\zeta'^2 \label{cz2case}
 \ee
and so
\be m^2_{\zeta'} = \frac{(1+16 \beta^2/\bp^4)}{48 c}\bp^2 > 0. 
\label{mzp22}\ee
In this simple quadratic case, by using the expression of $K$ in~(\ref{cz2case}) one obtains
\be \omega(\zeta')= \sqrt{\frac{3}{2}} \bp \left[\tanh ^{-1}\left(\frac{4 B(\zeta')}{\sqrt{1+16 B(\zeta')^2}}\right)-\tanh ^{-1}\left(\frac{4 \beta }{\sqrt{\bp^4+16 \beta ^2}}\right)\right].\ee
By inverting this function one then finds
\be  \zeta'(\omega) = \frac1{2c}\left(\frac{\bp^2 \tanh X(\omega)}{4\sqrt{1-\tanh^2X(\omega)}}-\beta\right),\label{zetapomega}\ee
where
\be X(\omega)\equiv \sqrt{\frac{2}{3}}\frac{\omega}{\bp}+\tanh ^{-1}\left(\frac{4 \beta }{\sqrt{16 \beta ^2+\bp^4}}\right) \label{Xofomega}  \ee
and the potential is 
\be U(\zeta'(\omega)) = c\zeta'(\omega)^2=\frac1{4c}\left(\frac{\bp^2 \tanh X(\omega)}{4\sqrt{1-\tanh^2X(\omega)}}-\beta\right)^2. \label{Uofomega}\ee
We see that the condition $c>0$, which ensures $m^2_{\zeta'}>0$, also ensures that the potential is bounded from below.
 The function $\zeta'(\omega)$ at large field values is (using $(1-\tanh^2(x))\exp(2x)\to 4$ as $x\to\infty$) 
\be \zeta' (\omega)=  \frac{\bp^2}{16 c} \text{sign}(\omega)\exp\left(\sqrt{\frac{2}{3}}\frac{|\omega|}{\bp}\right), \quad (|\omega|\gg\bp)\ee
 and one obtains an exponential potential:  
 \be U(\zeta'(\omega)) = \frac{\bp^4}{256 c} \exp\left(\sqrt{\frac{8}{3}}\frac{|\omega|}{\bp}\right), \quad (|\omega|\gg\bp).\ee
 On the other hand, at small field values
 \be \zeta' (\omega)= \frac{m_{\omega}\omega}{\sqrt{2c}}, \quad U(\zeta'(\omega)) = \frac{m_{\omega}^2\omega^2}{2} \quad (|\omega|\ll\bp),\ee
 where $m_{\omega} = m_{\zeta'}$.
 For intermediate values of $\omega$ the potential is shown in Fig.~\ref{potzetap}. We note that the behavior in the intermediate region, unlike the one at large field values, depends crucially on the Barbero-Immirzi parameter.
 The plots also show the invariance of the potential under $\{\omega, \beta\}\to \{-\omega, -\beta\}$ which can be analytically understood from Eqs.~(\ref{Xofomega}) and~(\ref{Uofomega}).

\subsubsection{Dynamical combination of $\zeta$ and $\zeta'$}

In general, for actions of the form~(\ref{Seq-DN}) a combination of  $\zeta$ and $\zeta'$ can be dynamical. Although a dynamical combination of  $\zeta$ and $\zeta'$ is not an unambiguous sign of dynamical distorsion (as $\zeta$ is sourced not only by the distorsion, but by the metric too,  see Eq.~(\ref{RRC})), here we explicitly compute the kinetic and potential terms of such dynamical combination in simple and illuminating cases. We do so in order to compare them with the most interesting case where the distorsion field $\zeta'$ is clearly dynamical, which we have analyzed in Sec.~\ref{Dynamicalzp}.

\vspace{0.2cm}

As an example, we first consider  the case where $\Delta$ depends on ${\cal R}$ and ${\cal R'}$ only through a combination $a(\Phi) {\cal R} + b(\Phi)  {\cal R'}$ that is linearly independent of $\alpha(\Phi) {\cal R} + \beta(\Phi)  {\cal R'}$. This   linear independence is assumed in order not to fall into the cases examined in Sec.~\ref{fake}, which have been proved not to contain extra degrees of freedom besides the metric, and the matter fields $\{\phi, \psi, A_\mu^I\}$.
Let us assume for simplicity again that these  matter fields are absent so that
\bea S &=&   \int d^4x\sqrt{-g}(\alpha {\cal R} + \beta{\cal R'} + \Delta(a  {\cal R} + b   {\cal R'})) \nonumber \\
&& =  \int d^4x\sqrt{-g}\left[\left(\alpha+a\frac{\partial\Delta}{\partial z}(z)\right) {\cal R} + \left(\beta+b\frac{\partial\Delta}{\partial z}(z)\right){\cal R'} + \Delta(z) -z \frac{\partial\Delta}{\partial z}(z)\right] , \label{Szeta(p)}\eea
where in the second step we introduced   the auxiliary field $z$ and we assumed, again without loss of generality, $\frac{\partial
^2\Delta}{\partial z^2}(z)\neq 0$.  
The two functions in front of ${\cal R}$ and ${\cal R'}$ can only be proportional to each other when $\frac{\partial\Delta}{\partial z}$ is constant (which is not compatible with $\frac{\partial
^2\Delta}{\partial z^2}(z)\neq 0$) and/or when $\{a,b\}$ and $\{\alpha,\beta\}$ are linearly dependent (which has been excluded in this case). So it is not possible to remove both functions with a rescaling of the metric $g_{\mu\nu}\to \Omega^2 g_{\mu\nu}$. We can, however, convert the function in front of ${\cal R}$ into $\bp^2/2$ by choosing
\be \Omega^2(z) = \frac{\bp^2}{2(\alpha+a\frac{\partial\Delta}{\partial z}(z))}, \ee  
whenever $\alpha+a\frac{\partial\Delta}{\partial z}(z)> 0$, which we assume from now on. After this metric rescaling
\be S =   \int d^4x\sqrt{-g}\left[\frac{\bp^2}{2}{\cal R} +\bp^2B(z){\cal R'} - U(z)\right], \ee
where
\be B(z) = \frac{\beta + b\frac{\partial\Delta}{\partial z}(z)}{2(\alpha+a\frac{\partial\Delta}{\partial z}(z))}, \qquad U(z) = \frac{\bp^4(z\frac{\partial\Delta}{\partial z}(z)-\Delta(z))}{4(\alpha+a\frac{\partial\Delta}{\partial z}(z))^2}. \label{BUofz}\ee
 By using again~(\ref{RRC}) and~(\ref{RpRC}) and integrating out $C_{\mu~\sigma}^{~\,\rho}$ as we did in Sec.~\ref{Dynamicalzp} we obtain
 \be S = \int d^4x\sqrt{-g}\left[\frac{M_P^2}{2} R-K(z)\frac{(\partial z)^2}{2} -U(z) \right], \label{SeqzpD2}\ee
 where
 \be K(z) = \frac{24\bp^2}{1+16 B^2(z)}\left(\frac{\partial B}{\partial z}\right)^2. \label{Kcomb1}\ee
It is easy to show that $\frac{\partial B}{\partial z}\neq 0$ when $\frac{\partial
^2\Delta}{\partial z^2}(z)\neq 0$ and $\{a,b\}$ and $\{\alpha,\beta\}$ are linearly independent.  So $z$ has a non-vanishing kinetic term and is thus a dynamical field in this case. Also, $K(z)$ is always positive, so $z$ is never a ghost. Like we did before, we can render the kinetic term of this dynamical scalar canonical through the redefinition $\omega(z)$ in~(\ref{omegazeta}) and express the action in terms of $\omega$ like we did in~(\ref{SeqCan}).
 
 Let us determine now the mass of $z$ (defined as the mass of the fluctuations of $z$ around a Lorentz invariant solution). By construction on a solution of the field equation $z = a {\cal R}+b {\cal R'}$, as it can be easily checked from~(\ref{Szeta(p)}), so in a Lorentz invariant stationary point $z=0$ (see Eqs.~(\ref{RRC}) and~(\ref{RpRC})). Note that Lorentz invariance also requires $U(0)=0$ and so, using the second expression in~(\ref{BUofz}), also $\Delta(0)=0$ and \be\frac{\partial U}{\partial z}(0)=0\qquad \frac{\partial
^2U}{\partial z^2}(0)=\frac{\bp^4 \frac{\partial
^2\Delta}{\partial z^2}(0)}{4 \left(\alpha +a \frac{\partial \Delta}{\partial z}(0)\right)^2}.\ee Expanding the action in~(\ref{SeqzpD2}) at quadratic order in $z$ we then easily obtain the squared mass of $z$:
 \be m_z^2 = \frac{\bp^2(1+16 B^2(0))\frac{\partial^2 \Delta}{\partial z^2}(0)}{96\left(\alpha+a\frac{\partial \Delta}{\partial z}(0)\right)^2\left(\frac{\partial B}{\partial z}(0)\right)^2}. \label{mz2} \ee
 Given the assumption we have made, $m_z^2$ is always finite and non vanishing. It is also positive for $\frac{\partial^2 \Delta}{\partial z^2}(0)> 0$, which is then the condition in order for $z$ not to be a tachyon.
 
 The potential of $z$ can only be computed once we specify the function $\Delta$. As an example, we take now a quadratic function like we did in Sec.~\ref{Dynamicalzp},  $\Delta(z) = c z^2$, where $c$  is a  positive constant (so that $z$ is not a tachyon). In this case we obtain 
\bea && B(z) = \frac{\beta +2 b c z}{ 2\alpha +4 a c z},\quad U(z) =\frac{c \bp^4 z^2}{4 (\alpha +2 a c z)^2}, 
   \label{mz22}\\ &&\frac{\partial B}{\partial z} =\frac{c (\alpha  b-a \beta )}{(\alpha +2 a c z)^2}, \quad K(z) = \frac{24 c^2 \bp^2 (\alpha  b-a \beta )^2}{(\alpha+2 a c z )^4 \left[1+\frac{4 (\beta +2 b c z)^2}{(\alpha+2 a c z)^2}\right]},  \quad  m^2_z = \frac{\bp^2 \left(\alpha ^2+4 \beta ^2\right)}{48 c (\alpha  b-a \beta )^2} > 0. \nonumber
\eea
Note that the quantity $\alpha  b-a \beta$ never vanishes because $\{a,b\}$ and $\{\alpha,\beta\}$ have been assumed to be linearly independent. In this case the potential $U(z)$ is asymptotically flat at large $z$, unlike the $U(\zeta')$  considered in Sec.~\ref{Dynamicalzp} at large $\zeta'$. 
 However, expressing $z$ in terms of $B$ through the first equation in~(\ref{mz22}) to find $U$ as a function of $B$ we obtain
 \be U = \frac{\bp^4 (2 \alpha  B-\beta)^2}{16 c (\alpha  b-a \beta )^2},\ee 
 which is, surprisingly,  the same potential  as the one in~(\ref{cz2case}) once we express $\zeta'$ in terms of $B$ and we redefine the parameters appropriately. Given that the kinetic term of $B$ is also the same (see the first expression in~(\ref{KBUzetap}) and~(\ref{Kcomb1})) this scalar-tensor theory is precisely the same as the one of Sec.~\ref{Dynamicalzp}, which features a dynamical distorsion.

\vspace{0.4cm}

Let us consider now another example. A combination of  $\zeta$ and $\zeta'$ can be dynamical for actions of the form~(\ref{Seq-DN}) also when the Hessian matrix of $\Delta$ (with respect to ${\cal R}$ and ${\cal R'}$) is not singular.
This example, as we will see, is a bit more complicated to analyze, but it can be considered as a more generic case: $\Delta$ can be expected to depend on both ${\cal R}$ and ${\cal R'}$ rather than on a specific linear combination of them. To illustrate how a kinetic term   can emerge we take again the simple case where there are no matter fields $\{\phi, \psi, A_\mu^I\}$ so that, introducing the two auxiliary fields $\zeta$ and $\zeta'$, we can write
 \bea S &=& \int d^4x\sqrt{-g}\left[ \left(\alpha+\frac{\partial \Delta}{\partial \zeta}(\zeta,\zeta')\right){\cal R}+\left(\beta+\frac{\partial \Delta}{\partial \zeta'}(\zeta,\zeta')\right) {\cal R'}\right. \nonumber \\&&\left. + \Delta(\zeta,\zeta')-\zeta\frac{\partial \Delta}{\partial\zeta}(\zeta,\zeta')-\zeta'\frac{\partial \Delta}{\partial\zeta'}(\zeta,\zeta')\right].  \eea
 By performing again a local rescaling of the metric $g_{\mu\nu}\to \Omega^2 g_{\mu\nu}$ with 
 \be \Omega^2 = \frac{\bp^2}{2 \left(\alpha +\frac{\partial\Delta}{\partial\zeta}(\zeta,\zeta')\right)} \ee
 (having assumed $\alpha +\frac{\partial\Delta}{\partial\zeta}(\zeta,\zeta')>0$)
 we obtain
 \bea S &=& \int d^4x\sqrt{-g}\left[  \frac{\bp^2}{2} {\cal R}+  \bp^2 B(\zeta,\zeta') {\cal R'} -U(\zeta,\zeta')\right],  \eea
 where this time
 \be B(\zeta,\zeta') = \frac{\beta +\frac{\partial\Delta}{\partial\zeta'}(\zeta,\zeta')}{2(\alpha +\frac{\partial\Delta}{\partial\zeta}(\zeta,\zeta'))},\quad U(\zeta,\zeta')=\frac{\bp^4}{4\left(\alpha +\frac{\partial\Delta}{\partial\zeta}(\zeta,\zeta')\right)^2}\left(\zeta\frac{\partial \Delta}{\partial\zeta}(\zeta,\zeta')+\zeta'\frac{\partial \Delta}{\partial\zeta'}(\zeta,\zeta')- \Delta(\zeta,\zeta')\right). \nonumber \ee
 Note that $B$ generically depends on both $\zeta$ and $\zeta'$.
 By using~(\ref{RRC}) and (\ref{RpRC}) and integrating out $C_{\mu~\sigma}^{~\,\rho}$ as we did in Sec.~\ref{Dynamicalzp} we obtain  
 \be S = \int d^4x\sqrt{-g}\left\{\frac{M_P^2}{2} R-K(B(\zeta,\zeta'))\frac{(\partial B)^2}{2}-U(\zeta,\zeta') \right\}, \ee
 where
 \be K(B)= \frac{24\bp^2}{(1+16 B^2)}. \label{KofB3}\ee
 Therefore, the field $B(\zeta,\zeta')$ is the   dynamical combination of $\zeta$ and $\zeta'$. Since $K(B)$ is always positive, $B$ is never a ghost. 
 
 In order to compute the potential of $B$ we need to  integrate out the other independent combination of $\zeta$ and $\zeta'$ that is not dynamical. We can do so by imposing that $U$ is stationary with respect to variations of $\zeta$ and $\zeta'$ with constant $B(\zeta,\zeta')$. Calling $b$ such constant value, when $\zeta$ is varied $\zeta'$ must equal $\zeta'_b(\zeta)$, which is the function of $\zeta$ such that $B(\zeta, \zeta'_b(\zeta)) = b$. Assuming that $\zeta'_b(\zeta)$ is a single-valued differentiable function, the condition that $U$ is stationary with respect to variations of $\zeta$ and $\zeta'$ with constant $B(\zeta,\zeta')$ can be expressed as follows 
 \be\left.\frac{\partial U}{\partial\zeta}+ \frac{\partial U}{\partial\zeta'}\frac{d\zeta'_b}{d\zeta}\right|_{b=B(\zeta,\zeta')}=0. \label{stationarity}\ee   
 Imposing this constraint on $\zeta$ and $\zeta'$ integrates out the other non-dynamical scalar  and allows us to express $U$ in terms of $B$ only. The resulting action is
  \be S = \int d^4x\sqrt{-g}\left\{\frac{M_P^2}{2} R-K(B)\frac{(\partial B)^2}{2}-U(B) \right\}. \ee
  Once again, we can render the kinetic term of this dynamical scalar canonical through the redefinition $\omega(B)$ in~(\ref{omegazeta}) and express the action in terms of $\omega$ like we did in~(\ref{SeqCan}).
 
 We cannot determine explicitly $U(B)$ until we specify the function $\Delta$.  As an example let us consider the case where $\Delta$ is  a generic quadratic function of $\zeta$ and $\zeta'$, namely $$\Delta(\zeta,\zeta')=c \zeta^2+c'\zeta'^2+c_m \zeta\zeta',$$ whose Hessian matrix is not singular for $4 c c'\neq c_m^2$. In this case 
 \be B(\zeta,\zeta')=\frac{\beta +c_m \zeta +2 c' \zeta'}{2 (\alpha +2 c \zeta +c_m \zeta')}, \qquad U(\zeta,\zeta')=\frac{\bp^4 \left(c \zeta ^2+c' \zeta'^2+c_m\zeta\zeta'\right)}{4 (\alpha +2 c \zeta +c_m\zeta')^2}\ee
and one finds (for $b c_m\neq c'$)
 \bea\zeta'_b(\zeta) &=&\frac{\beta-2 \alpha  b+(c_m -4 b c)\zeta }{2 (b c_m-c')}, \\ 
 \left.\frac{\partial U}{\partial\zeta}+ \frac{\partial U}{\partial\zeta'}\frac{d\zeta'_b}{d\zeta}\right|_{b=B(\zeta,\zeta')} &=& \frac{\bp^4 \left(4 c c'-c_m^2\right) (\alpha  \zeta +\beta  \zeta')}{4 (\alpha +2 c \zeta +c_m \zeta')^2 [2 c' \alpha-c_m \beta+(4 cc'-c_m^2) \zeta ]}. \eea
Since the non singularity of the Hessian matrix of $\Delta$ requires $4 c c'\neq c_m^2$, integrating out the non-dynamical scalar through Eq.~(\ref{stationarity}) then gives $\alpha\zeta=-\beta\zeta'$. This condition, together with $B=B(\zeta,\zeta')$ allows us to express both $\zeta$ and $\zeta'$ in terms of $B$ and the potential of this dynamical scalar reads
 \be U(B) = \frac{\bp^4 (2 \alpha  B-\beta)^2}{16 \left[\beta  (\beta  c-\alpha  c_m)+\alpha ^2 c'\right]}.\label{UB3}\ee 
Again this is the same potential  as the one in~(\ref{cz2case}) once we express $\zeta'$ there in terms of $B$ and redefine the parameters appropriately. Like in the previous example, also the kinetic term of $B$ is the same (see the first expression in~(\ref{KBUzetap}) and~(\ref{KofB3})) so this scalar-tensor theory is again precisely the same as the one of Sec.~\ref{Dynamicalzp}, that features a dynamical distorsion. We then see that this theory is much more general than what we could have imagined from the analysis of Sec.~\ref{Dynamicalzp}.

 \vspace{0.4cm}
 
 One can of course find cases where both $\zeta$ and $\zeta'$ are dynamical: for example one can introduce, like in~(\ref{Seq}), a dependence of $\Sigma$ on $C$, which is not invariant but transforms inhomogeneously under $g_{\mu\nu}\to \Omega^2g_{\mu\nu}$ for a spacetime-dependent $\Omega$. But, as observed before, it is only $\zeta'$ that is directly linked to the distorsion. Since we are interested in a dynamical distorsion we do not explore these further possibilities here and leave them for future work.

 \subsection{Examples: Poincar\'e gauge theories coupled to matter}
\label{Poincare gauge theories coupled to matter}
 
 The distorsion, in the most general case, does not  only include scalars and pseudoscalars, but also higher rank tensors, which, in the most general case also lead to spin-3, spin-2 and spin-1 particles (see Ref.~\cite{Baldazzi:2021kaf} for a detail discussion and a summary of previous works). Here we consider the case of Poincar\'e gauge theories, also known as Einstein-Cartan theories (see~\cite{Hehl:1976kj,Shapiro:2001rz} for detailed reviews): the gravitational fields are represented by the tetrads and the connection, which, as we have seen in Sec.~\ref{Ingredients},  has  to be metric compatible, i.e. ${\cal D}_\rho g_{\mu\nu} =0$. From the physical point of view this is not a restrictive choice because, as we have seen in section~\ref{Ingredients}, in order to have fermions it is necessary to introduce the tetrads and have a metric-compatible connection. In this case the distorsion coincides with what is known as the contorsion, which can be expressed in terms of the torsion:
 \be C_{\mu\nu\rho} = \frac12 (T_{\mu\nu\rho}+T_{\nu\mu\rho}-T_{\mu\rho\nu}),\ee
 which is antisymmetric in the second and third indices.
 From this equation and~(\ref{torsion-distorsion}) we see that the contorsion vanishes if and only if the torsion does.
  As we have seen in Sec.~\ref{Ingredients}, the tetrads are defined modulo local Lorentz transformations, which together with the local translations always present in any generally covariant theory, leads to local Poncar\'e symmetry (hence the name Poincar\'e gauge theories).

 As shown in~\cite{Neville:1978bk} (see also Refs.~\cite{Percacci:2020ddy,Neville:1979rb} for subsequent studies), in Poincar\'e gauge theories  in the absence of matter fields (i.e. without $\{ \phi, \psi, A_{\mu}^I \}$) the metric and the connection generically contain three spin-2 fields (one of which correspond to the ordinary massless graviton), plus four spin-1 and three spin-0 fields (including the fields $\zeta$ and $\zeta'$ discussed in Sec.~\ref{(pseudo)scalaron}). The spin-3 field present in the most general case is removed by the condition of metric compatibility. Subsequently, it was shown  that the stability of these theories can only occur if the additional spin-2 fields (besides the ordinary graviton) are massive at least in the absence of matter fields~\cite{Neville:1981be}. The argument was based on an expansion of the action at the quadratic level in the fluctuations around the flat (Minkowski) spacetime.

 If one introduces ordinary matter fields $\{\phi, \psi, A_\mu^I\}$ this result does not change as we now show. To see this let us first introduce some scalar or pseudoscalar fields $\phi$. Since we want to exclude the presence of massless spin-2 fields we take these scalars to be massless because otherwise it would not be possible to construct a quadratic mixing term between them and the massless components of the contorsion. The only possible independent scalar or pseudoscalar terms involving the contorsion and $\phi$ at the quadratic level and with only one derivative  are then
 \be C_{\mu\nu}^{~~~\mu}\partial^\nu\phi, \qquad 
\epsilon^{\mu\nu\rho\sigma}  C_{\mu\nu\rho}\partial_\sigma\phi, \label{Cphi}
 \ee 
 which, of course, can only be constructed with those $\phi$ fields that are invariant under the gauge group $G$. 
The terms in~(\ref{Cphi}) are mixing terms between $\phi$ and a vector field $C_{\mu\nu}^{~~~\mu}$ and a pseudovector field $\epsilon^{\mu\nu\rho\sigma}  C_{\mu\nu\rho}$. So they do not affect the spin-2 sector. Actually the quadratic terms in~(\ref{Cphi}) even vanish in the massless sector as one can always decompose the above mentioned vector and pseudovector fields into spin-1 fields that are transverse and spin-0 fields whose d'Alembertian is anyhow zero in the massless case. Non-vanishing scalar or pseudoscalar terms with more than one derivative cannot be constructed either as they would unavoidably contain (because $C_{\mu\nu\rho}$ is antisymmetric in the second and third indices) a d'Alembertian acting on $\phi$, which vanishes because $\phi$ are massless fields.

 Similarly, considering gauge fields, one can construct quadratic terms that involve both the contorsion and an Abelian gauge field $A_\mu$, such as
 \be C_{\mu\nu\rho}\partial^\mu F^{\nu\rho}, \quad C_{\nu\mu\rho} \partial^\mu F^{\nu\rho},\quad C_{\alpha\nu}^{~~~\alpha} \partial_\mu F^{\mu\nu}, \quad \epsilon^{\mu\nu\rho\sigma}C_{\mu\nu\rho} \partial_\alpha F^\alpha_{~~\sigma}, \quad  \epsilon^{\mu\nu\rho\sigma} C_{\mu\nu\alpha} \partial^\alpha F_{\rho\sigma}, \quad  ... \,  ,  \label{spin1Mod} \ee
  where $F_{\mu\nu}$ is the field strength of $A_\mu$.
But it is always possible to choose the gauge in a way that fields with a non-vanishing spin are described by transverse tensors so, recalling that the d'Alembertian of any massless field vanishes, these terms do not modify the spin-2 sector. Of course, with fermions it is not possible to construct terms involving  $C_{\mu\nu\alpha}$  that change the quadratic action because fermions always come in pair.
 
We conclude that, even in the presence of matter fields, the argument of~\cite{Neville:1981be} holds and the two extra spin-2 fields besides the ordinary graviton must be massive to have a stable theory.

\subsubsection{Dark photons from torsion}\label{Dark photons from torsion}
 The vector $v_\nu\equiv C_{\alpha\nu}^{~~~\alpha}$,
and the pseudovector $p^\sigma\equiv \frac{\epsilon^{\mu\nu\rho\sigma}}{\sqrt{-g}} C_{\mu\nu\rho}$, that we have already discussed in the previous section, contain spin-1 particles, which  can play the role of dark photons of gravitational origin. Dark photons have  interesting phenomenology (see e.g.~\cite{DP,Belyaev:2007fn}) as they can act as portals to dark sectors.

Note that after integrating by parts the third and fourth terms in~(\ref{spin1Mod}) one obtains mixing kinetic  terms between the vector $v_\mu$ and an Abelian gauge field $A_\mu$ and between
the pseudovector $p_\mu$ and $A_\mu$,
\be  v_{\mu\nu} F^{\mu\nu}, \quad p_{\mu\nu} F^{\mu\nu}, \label{mixingDP}\ee
where
\be v_{\mu\nu}\equiv \partial_\mu v_\nu- \partial_\nu v_\mu, \qquad p_{\mu\nu}\equiv \partial_\mu p_\nu- \partial_\nu p_\mu\ee
are the field strengths of $v_\mu$ and $p_\mu$. If $F_{\mu\nu}$ is the electromagnetic field strength the terms~(\ref{mixingDP}) are mixing terms between the photon and the torsion dark photons. These mixing terms give the possibility of detecting the effect of the dark photons when they are massive~\cite{DP}. In the massless case interaction terms between the dark photons and the SM fields are necessarily higher dimensional (non-renormalizable) operators~\cite{Dobrescu:2004wz} that might, however, induce observable effects depending on the size of their coefficients. Such higher dimensional operators are allowed in our EFT approach (generically, the couplings of $v_\mu$ and $p_\mu$ in the metric theory depends on the initial metric-affine action~\cite{Diakonov:2011fs}). 

One sees that theories where the connection carries extra degrees of  freedom (besides the metric) generically lead to the existence of (and thus motivate) dark photons. In total there are two dark photons with negative parity and two with positive parity: $v_\mu$, $p_\mu$ and other two spin-1 fields (one with positive parity and another one with negative parity) that come from the other independent components of the torsion, as it can be easily shown by using the results of~\cite{Neville:1978bk}. 
 
 One might think that the torsion spin-1 fields  cannot couple to the (pseudo)scalars $\phi$ because the torsion is part of the full connection (and (pseudo)scalars are invariant under proper orthochronous Poincar\'e transformations). However, in the most general Poincar\'e gauge theory we could also include these spin-1 fields in the covariant derivative of $\phi$ by adding to the action appropriate terms: considering, as an example, $v_\mu$ such a  term would be
 \bea && \int d^4x\sqrt{-g}\left(-\left[({\cal D}_\mu+iv_\mu) \phi \right]^\dagger({\cal D}^\mu+iv^\mu) \phi +{\cal D}_\mu\phi^\dagger {\cal D}^\mu\phi\right)\nonumber\\ 
 &&\int d^4x\sqrt{-g}\left(iv_\mu \left[\phi^\dagger{\cal D}^\mu\phi-\left({\cal D}^\mu  \phi \right)^\dagger\phi \right]-v_\mu v^\mu\phi^\dagger\phi\right)
 \eea
 and analogous terms for the other spin-1 fields.
 It is clear that these terms depend on $\phi$, ${\cal D}_\mu\phi$ and  $C_{\mu~\sigma}^{~\,\rho}$ and can, therefore, be included in a function like $\Sigma(\Phi,{\cal D}\Phi, C)$ in Eqs.~(\ref{Seq1}) and~(\ref{Seq}).
 
 \vspace{0.4cm}
 
 We do not study here cases where the torsion spin-2 fields are dynamical due to standard difficulties when one attempts an extension to a fully covariant theory in the presence of additional spin-2 fields besides the graviton, see e.g. Ref.~\cite{Neville:1981be}.

 \subsubsection{Coupling the pseudoscalaron to matter}
\label{Coupling the pseudoscalaron to matter}
     
     One of the most interesting component of the distorsion, that can be dynamical, is the pseudoscalaron $\zeta'$, which we have discussed in Sec.~\ref{(pseudo)scalaron}. This field is also present in Poincar\'e gauge theories, because in Sec.~\ref{(pseudo)scalaron} we have not used that ${\cal D}_\rho g_{\mu\nu}\neq 0$.

In order to illustrate how the pseudoscalaron couples to a generic matter sector let us take an action of the form
 \be S = \int d^4x\sqrt{-g}\left[\alpha(\phi){\cal R}+\beta(\phi){\cal R'} + \Delta(\phi,{\cal R'})+\Sigma(\Phi,{\cal D}\Phi)\right], \label{Spsm}\ee
 where $\alpha$, $\beta$ and $\Delta$ are generic functions of the (pseudo)scalars $\phi$, the function $\Delta$ has an additional dependence on ${\cal R'}$, which has been added to introduce the pseudoscalaron (see Sec.~\ref{Dynamicalzp}), and 
\bea  
\Sigma(\Phi,{\cal D}\Phi) &=&  
 -\frac{{\cal D}_\mu \phi_k \, {\cal D}^\mu \phi_k}{2} - V(\phi)  - \frac14 F_{\mu\nu}^I F^{\mu\nu I}\nonumber\\ &&+\frac12 (\bar\psi_j i\slashed{\cal D} \psi_j  -M_{ij}\psi_i\psi_j - Y^k_{ij} \psi_i\psi_j \phi_k + \hbox{h.c.}), \label{MLag}
\eea 
represents the matter Lagrangian, where 
$V$ is the potential. The coefficients $Y^a_{ij}$  and $M_{ij}$ are generic Yukawa couplings and fermion mass parameters.
  As usual, since we work with Weyl fermions, $\slashed{\cal D}\psi_j = \bar\sigma^\mu {\cal D}_\mu\psi_j$, where $\bar\sigma^\mu = e^\mu_a \bar\sigma^a$ and $\bar \psi$ represent the (transpose) hermitian conjugate of $\psi$.
All terms are contracted in a gauge-invariant way with respect to both the local Poincar\'e group and the gauge group $G$. This action is clearly a particular case of~(\ref{Seq-DN}). This form, despite not being the most general one, is suggested by the structure of the SM (although it also covers, among others, any of its renormalizable extensions) and by the geometrical interpretation of the torsion as part of the full connection: in~(\ref{Spsm}) we only use the covariant derivative ${\cal D}$ rather than the one, $D$, constructed with the Levi-Civita connection or, in other words, $\Sigma$ does not explicitly depend on the contorsion $C_{\mu~\sigma}^{~\,\rho}$. The matter Lagrangian in~(\ref{MLag}) is general enough to accomodate not only all the SM fields but also additional fields needed to describe the current evidence of beyond-the-SM physics (neutrino masses and mixings, dark matter, baryon asymmetry, etc.). 
 
 By performing steps similar to those made around Eq.~(\ref{Szetap}), the action in~(\ref{Spsm}) can be equivalently rewritten as follows
\be S = \int d^4x\sqrt{-g}\left[ \alpha(\phi){\cal R}+\bp^2B(\phi,\zeta'){\cal R'} + \Delta(\phi,\zeta') -\zeta'\frac{\partial \Delta}{\partial\zeta'}(\phi,\zeta')+\Sigma(\Phi,{\cal D}\Phi)\right]  \label{Szetap2}\ee
having required again, without loss of generality, $\frac{\partial^2\Delta}{\partial \zeta'^2}\neq 0$. Here the function $B(\phi,\zeta')$ is
\be B(\phi,\zeta')=\frac{\beta(\phi)+\frac{\partial \Delta}{\partial\zeta'}(\phi,\zeta')}{\bp^2}. \ee 

In~(\ref{Szetap2}) the pseudoscalaron $\zeta'$ appears explicitly, but the other torsion components  are not dynamical like in Sec.~\ref{Dynamicalzp}. We can again integrate out the torsion by using the method of Appendix~\ref{appendixInt} to find 
\bea S&=& \int d^4x\sqrt{-g}\left[\alpha(\phi)R - \frac14 F_{\mu\nu}^I F^{\mu\nu I} - \frac{D_\mu \phi_k \, D^\mu \phi_k}{2} - U(\phi,\zeta') \right. \nonumber\\
&&\hspace{2.2cm} +\frac12 ( \bar\psi_j i\slashed{D} \psi_j-M_{ij}\psi_i\psi_j -  Y^k_{ij} \psi_i\psi_j \phi_k + \hbox{h.c.})\nonumber\\
&&\hspace{2.2cm} \left.-\frac{\alpha(\phi)V_\mu V^\mu-\frac{\alpha(\phi)}{4}\partial_\mu\alpha(\phi)\partial^\mu\alpha(\phi)-2\bp^2B(\phi,\zeta')\partial_\mu\alpha(\phi)V^\mu}{\frac23\bp^4(B^2(\phi,\zeta')+\frac{\alpha(\phi)^2}{4\bp^4})}\right], \label{SintJ}\eea
 where the full potential is 
\be U(\phi,\zeta') = V(\phi)-\Delta(\phi,\zeta')+\zeta' \frac{\partial \Delta}{\partial\zeta'}(\phi,\zeta')\ee
and $V_\mu$ is defined by
\be V_\mu \equiv \bp^2\partial_\mu B(\phi,\zeta') +\frac18 \bar\psi_j \bar \sigma_\mu  \psi_j.  \ee
Note that $U(\phi,\zeta')$ contains some interactions of $\zeta'$ with the $\phi$ fields, e.g. the Higgs. The last line in Eq.~(\ref{SintJ}) contains other interactions of $\zeta'$ as well as its kinetic term, which emerges from the $V_\mu V^\mu$ term. Note that in this class of theories the pseudoscalaron interacts with   $\phi$ and the fermions $\psi$, but not with the gauge fields $A^I_\mu$: this is because the starting action~(\ref{Spsm}) does not feature couplings between the torsion and $A^I_\mu$. The pseudoscalaron here interacts with $\phi$ through the function $\alpha$ and the potential $U$ and also has two- and four-fermion interactions.

A commonly encountered  case is $\alpha(\phi) = \bp^2/2+\xi_{kl}\phi_k\phi_l$, where $\xi_{kl}$ are real coefficients, sometimes called non-minimal  couplings. In this case one recovers the standard Einstein-Hilbert action for gravity at small field values, when $\alpha \simeq \bp^2/2$. One can easily compute the interactions in terms of the  $\xi_{kl}$ (including those involving the pseudoscalaron) by expanding $\alpha(\phi)$ in powers of $\xi_{kl}\phi_k\phi_l/\bp^2$.

 \section{A note on the equivalence principle} \label{A note on the equivalence principle} 

In any modification or extension of GR   it is natural to ask whether (and to what extent) the equivalence principle holds. It is particularly interesting to answer this question in the context of metric-affine theories as these are gravitational theories constructed starting from the geometrical principle of general covariance.

Let us first recall what the equivalence principle states: {\it for any fixed spacetime point $X$,  it is possible to choose a reference frame (called locally inertial frame) where  the laws of physics are those without gravity in a small enough  neighbourhood of $X$.} 

A first thing one may note is that the equivalence principle is ambiguous if one does not specify what     is meant by ``the laws of physics without gravity". In order to eliminate this ambiguity, given our current description of fundamental non-gravitational forces, we understand that the physics without gravity is described by a theory with ordinary matter, such as the one present in the SM and its common extensions. This can feature (pseudo)scalars, gauge fields  and  fermions, which are enough to account for all matter we observe and address the  evidence of beyond-the-SM physics. Massive (pseudo)vector fields, for example, can be modeled by gauge fields and (pseudo)scalars using the St\"uckelberg or Higgs mechanism. Also note that   pseudoscalars and pseudovectors are  present in the QCD spectrum and appear in popular SM extensions, such as those featuring an axion. 
Therefore, the scalar $\zeta$ and pseudoscalar $\zeta'$, which we defined in Sec.~\ref{(pseudo)scalaron}, as well as the vector $v_\mu$ and pseudovector $p_\mu$ encountered in Sec.~\ref{Dark photons from torsion} are particular examples of ordinary matter fields. Since, starting from general covariance, gravity is described by the metric $g_{\mu\nu}$ and the connection ${\cal A}_{\mu~\sigma}^{~\,\rho}$, as discussed in Sec.~\ref{Ingredients}, we conclude that the equivalence principle tells us that 
{\it in the locally inertial frame} $g_{\mu\nu}(X)=\eta_{\mu\nu}$ and the effect of ${\cal A}_{\mu~\sigma}^{~\,\rho}(X)$ is indistinguishable from that of such ordinary matter.

 Another part of  the equivalence principle that calls for a clarification are the words ``small enough". Following the argument in~\cite{WeinbergGravity}, we interpret them as the requirement that not only $g_{\mu\nu}(X)=\eta_{\mu\nu}$, but also $\partial_\rho g_{\mu\nu}(X) = 0$ {\it in the locally inertial frame}. With this interpretation the equivalence principle also tells us that  $ \Gamma_{\mu~\sigma}^{~\,\rho}(X) =0$ and the effect of   $C_{\mu~\sigma}^{~\,\rho}(X)$ is indistinguishable from that of ordinary matter in the locally inertial frame.
So any physical effect of this $C_{\mu~\sigma}^{~\,\rho}(X)$ that cannot be accounted for by ordinary matter may be interpreted as a violation of the equivalence principle (see also Ref.~\cite{VonDerH} for a related discussion).

 It is important to note that a violation of this principle  can even occur in a metric theory, through the presence of higher dimensional terms in the action, which start to be relevant at high energies. An example is the term  $\int d^4x \sqrt{-g}\,  R F_{\mu\nu}^IF^{I\mu\nu}$: in a spacetime where $R\neq 0$ locally, such as the de Sitter spacetime of cosmological relevance, this term would lead to an observable modification of  electrodynamics due to gravity even in arbitrarily small neighbourhood of $X$. This is not surprising because the equivalence principle is a classical local statement but at very small distances, i.e. at very high energies, we expect quantum gravity effects to show up and these can lead to higher dimensional terms in the EFT description, such as the one we have just mentioned. The (classical) equivalence principle is expected to fail in a quantum gravity framework, while general covariance can survive~\cite{Bjerrum-Bohr:2015vda}.

On the other hand, as we have seen in Sec.~\ref{Poincare gauge theories coupled to matter},  starting from the general relativity principle,  the dynamical components of the distorsion that can be massless are only spin-1 and spin-0 fields for realistic theories (that must be stable and feature fermions and whose connection is, therefore, metric compatible). So at low enough energies the effect of $C_{\mu~\sigma}^{~\,\rho}$ is indistinguishable from that of ordinary matter not only at $X$ in the locally inertial frame, but in any frame and at any point. Furthermore,  in the low energy limit metric-affine theories coupled to spin-0, spin-1/2 and spin-1 fields are described by the Einstein-Hilbert term computed with the LC connection, Eq.~(\ref{EHt}), plus the renormalizable action of the matter fields $\{ \phi', \psi, A_{\mu}^{'I} \}$ 
(where $\phi'$ and $A_{\mu}^{'I}$ include $\phi$ and $A_{\mu}^I$ plus all spin-0 and spin-1 massless dynamical fields from the torsion), which do satisfy the equivalence principle. This result does not change if one also considers other fields with spin 3/2 or higher than or equal to two: the only massless particles with spin higher than or equal to two that can interact with gravity in a Minkowski background are gravitons and massless spin 3/2 particles should interact exactly as gravitinos in supergravity~\cite{Benincasa:2007xk,Porrati:2008rm}. But supersymmetry must be broken at low energies in order for the theory to be realistic and as soon as this happens the gravitino acquires a mass. 

Therefore, we see that, although general covariance does not imply the equivalence principle at all energies, the latter in general emerges at low energies from the former in realistic theories.

 \section{Conclusions}\label{Conclusions}

 We conclude by providing a detailed summary of the new results of this paper with some further discussions.
  \begin{itemize}
  \item After an introduction and  some background material in Secs.~\ref{Introduction} and~\ref{Ingredients}, in Sec.~\ref{eq-theories} we have constructed the most general action  of metric-affine EFTs that are equivalent to metric ones, namely those theories with a non-dynamical distorsion. We have included a generic matter sector featuring an arbitrary number of spin-1, spin-1/2 and spin-0 fields. We have pointed out, however, that in some specific cases the action can be brought in that form  with appropriate redefinitions although it might not look so initially. The bottom line of that section is that the actions with non-dynamical distorsion are those that can be recast in a form linear in the curvature  of the full connection  with the ``coefficients" of the linear terms  being independent of the distorsion itself. 
  This class is very vast and includes as a particular case, among many others, $f({\cal R})$ theories.
  \item In Sec.~\ref{Dyn-Dist} we have studied some examples of theories that have instead a dynamical distorsion. 
  
  We have investigated in detail a vast class where the parity-odd Holst invariant ${\cal R}'$ is a dynamical pseudoscalar field (pseudoscalaron). This field is supported by the distorsion (it vanishes when the distorsion does) and can, therefore, be regarded as a genuine distorsion field. The pseudoscalaron can coexist with a dynamical scalaron ${\cal R}$ and a generic matter sector. In the simplest cases we have been able to compute explicitly the pseudoscalaron kinetic term, mass and potential.

   In the same section, we have also discussed general Poincar\'e gauge theories coupled to matter, where the connection is metric compatible and fermions can be introduced. We have extended a previous result by Neville in a pure gravitational theory~\cite{Neville:1981be} to the presence of a generic matter sector, showing that the spin-2 fields from the torsion cannot be massless compatibly with the stability requirements and thus cannot appear at low enough energies. Also, we have commented on the possible phenomenology of torsion spin-1 fields, which can play the role of dark photons. At the end of Sec.~\ref{Dyn-Dist} we have computed interactions of the pseudoscalaron with a generic matter sector  and, of course, the metric. 
   
   These results can be used  in the future to study the role of the pseudoscalaron in the early and late universe as well as the possible scattering, production mechanisms and decays of this torsion field.
   \item Sec.~\ref{A note on the equivalence principle} presents a proof that in generic realistic,  and thus metric compatible, metric-affine EFTs the equivalence principle (appropriately defined) always emerges at low energies, although it is generically violated at high energies. This was possible by means of the extension of Neville's result to a general matter sector, which we presented in Sec.~\ref{Poincare gauge theories coupled to matter}: the massless dynamical torsion fields can only have spin 1 or spin 0 and can, therefore, be represented by ordinary matter fields; so at low enough energies the theory can be described by the Einstein-Hilbert action complemented by minimally-coupled ordinary matter fields, which satisfy the equivalence principle.
   \end{itemize}
 
 \subsection*{Acknowledgments}
 We thank Emanuele Orazi, Raffaele Savelli and Ilya Shapiro for useful discussions and Roberto Percacci for useful comments on the preprint. This work has been partially supported by the grant DyConn from the University of Rome Tor Vergata.
 
\appendix
\section{Integrating out the distorsion}\label{appendixInt}
Let us discuss here how the distorsion can be integrated out (i.e. how to determine the effective action after solving the distorsion field equations) for actions of the form~(\ref{Szetap2}). In the case where the connection is metric compatible, i.e. ${\cal D}_\rho g_{\mu\nu} =0$, this procedure has been performed in~\cite{Karananas:2021zkl} and~\cite{BeltranJimenez:2019hrm} and we have explicitly checked their results. 

We here show that, starting from\footnote{If one changes the starting action in a way that a metric compatible connection is no longer a solution of the connection field equations, like in e.g.~\cite{Rigouzzo:2022yan},  obviously one cannot show the same.}~(\ref{Szetap2}), the action of the effective metric theory that is obtained by integrating out the distorsion is the same even if the connection is  not necessarily metric compatible.  To this purpose we   actually demonstrate something more: if the field equations of the distorsion admit more than one solution  the effective action obtained by substituting the distorsion with {\it any} solution of its field equations is uniquely determined (i.e. such effective action does not depend on which solution for the distorsion we choose). 

After using~(\ref{RRC})-(\ref{RpRC}) and performing some integrations by parts, Lagrangians of the form~(\ref{Szetap2}) can be written as follows:
\be \frac12 x_i A_{ij}x_j-J_i x_i+Q, \label{distx}\ee
where the $x_i$ represent the components of the distorsion and $A_{ij}$, $J_i$ and $Q$ are real distorsion-independent coefficients, which can depend, however, on the other fields (the metric and the matter fields). We take $A_{ij}=A_{ji}$ without loss of generality. In the matrix formalism~(\ref{distx}) reads
\be \frac12 x^T Ax-J^T x+Q \label{distx2}\ee 
and the field equations of the distorsion are then 
\be Ax=J. \label{AxJ}\ee
This is a standard linear inhomogeneous equation with $A$ and $J$ real and $A^T=A$. If there are eigenvectors $x^{(n)}$ of $A$ with zero eigenvalues, $Ax^{(n)}=0$, there are solutions of Eq.~(\ref{AxJ}) if and only if $J^T x^{(n)}=0$ for all $n$. On the other hand, if the $x^{(n)}$ do not exist, i.e. $\det A\neq0$, there are no conditions on $J$ for the existence of solutions. Let us assume now, as we have already mentioned, that there are solutions of~(\ref{AxJ}), so that $J^T x^{(n)}=0$ for all $n$ if some $x^{(n)}$ exist. The general solution of~(\ref{AxJ}) can then be written 
\be x =A^{-1} J+ \sum_n c_n x^{(n)}, \ee
where the $c_n$ are arbitrary real coefficients that label all possible solutions.
Note that $A^{-1} J$ is well defined because $J$ is orthogonal to all $x^{(n)}$. We now plug this solution into the Lagrangian in~(\ref{distx2}) to obtain
\be Q-J^T \sum_n c_n x^{(n)}-\frac12 J^TA^{-1}J =Q-\frac12 J^TA^{-1}J, \ee
where we used $J^T x^{(n)}=0$ for all $n$. We see that if at least a solution of the field equations of the distorsion exists then the action is uniquely determined: this is because the dependence on the $c_n$ has disappeared.

 \vspace{1cm}
 
\footnotesize
\begin{multicols}{2}

\end{multicols}

\end{document}